\documentclass[useAMS,usenatbib]{mnras}

\usepackage[dvipsnames]{xcolor}
\usepackage{tikz,hyperref}

\definecolor{lime}{HTML}{A6CE39}
\DeclareRobustCommand{\orcidicon}{
	\begin{tikzpicture}
	\draw[lime, fill=lime] (0,0) 
	circle [radius=0.13] 
	node[white] {{\fontfamily{qag}\selectfont \tiny ID}};
	\draw[white, fill=white] (-0.0625,0.095) 
	circle [radius=0.007];
	\end{tikzpicture}
	\hspace{-2mm}
}

\foreach \x in {A, ..., Z}{\expandafter\xdef\csname orcid\x\endcsname{\noexpand\href{https://orcid.org/\csname orcidauthor\x\endcsname}
			{\noexpand\orcidicon}}
}
\usepackage{graphicx}
\usepackage{longtable}
\usepackage{amssymb,amsmath}
\usepackage{multicol}        % Multi-column entries in tables
\usepackage{bm}		% Bold maths symbols, including upright Greek
\usepackage{pdflscape}	% Landscape pages
\usepackage{array}
\usepackage{graphicx}

\newcommand{\heiiwr}{He\,{\sc ii}\,$\lambda4686$}

\newcommand{\ewhb}{EW(H$\beta$)}
\newcommand{\zsol}{Z$_{\odot}$}

\newcommand{\hii}{H\,{\small{\sc ii}}}
\newcommand{\ha}{H$\alpha$}
\newcommand{\hb}{H$\beta$}

\newcommand{\Av}{$\text{A}_{\text{V}}$}
\newcommand{\hei}{He\,{\sc i}}
\newcommand{\heii}{He\,{\sc ii}}

\newcommand{\oibyha}{[O\,{\sc i}]\,$\lambda6300$/H$\alpha$}

\newcommand{\sii}{[S\,{\sc ii}]}
\newcommand{\nia}{N\,{\sc i}}
\newcommand{\nii}{[N\,{\sc ii}]}
\newcommand{\niii}{N\,{\sc iii}}
\newcommand{\nv}{N\,{\sc v}}
\newcommand{\oi}{[O\,{\sc i}]}
\newcommand{\oii}{[O\,{\sc ii}]}
\newcommand{\oiia}{[O\,{\sc ii}]\,$\lambda3727$}
\newcommand{\oiii}{[O\,{\sc iii}]}
\newcommand{\siii}{[S\,{\sc iii}]}
\newcommand{\ciii}{C\,{\sc iii}}
\newcommand{\civ}{C\,{\sc iv}}
\newcommand{\ciiip}{C\,{\sc iii}]\,$\lambda1909$}
\newcommand{\ciiib}{C\,{\sc iii}\,$\lambda4636$}
\newcommand{\civr}{C\,{\sc iv}\,$\lambda\lambda5801,5812$}
\newcommand{\mgiia}{[Mg\,{\sc ii}]\,$\lambda2798$}

\newcommand{\neiiia}{[Ne\,{\sc iii}]\,$\lambda3869$}
\newcommand{\cliii}{[Cl\,{\sc iii}]}
\newcommand{\feiii}{[Fe\,{\sc iii}]}
\newcommand{\ariii}{[Ar\,{\sc iii}]}
\newcommand{\ariv}{[Ar\,{\sc iv}]}

\newcommand{\oiiibyhb}{[O\,{\sc iii}]\,$\lambda5007$/H$\beta$}
\newcommand{\niibyha}{[N\,{\sc ii}]\,$\lambda6583$/H$\alpha$}
\newcommand{\siibyha}{[S\,{\sc ii}]\,$\lambda6717+6731$/H$\alpha$}

\newcommand{\feiiia}{[Fe\,{\sc iii}]\,$\lambda4658$}

\newcommand{\ergs}{erg\,s$^{-1}$}

\newcommand{\logU}{$\log\langle U\rangle$}

\newcommand{\kms}{km\,s$^{-1}$}
\newcommand{\kmsMpc}{km\,s$^{-1}$\,Mpc$^{-1}$}
\title[L-S ring galaxy]{
Chemical abundances and ionizing mechanisms in the star-forming double-ring of AM\,0644-741 using MUSE data
}
\author[V.\,M.\,A. G\'omez-Gonz\'alez et al.]{V.\,M.\,A. G\'omez-Gonz\'alez\thanks{Email: vmagg@astro.physik.uni-potsdam.de}$^{1\orcidA{}}$, 
Y.\,D. Mayya$^{2\orcidB{}}$,
J. Zaragoza-Cardiel$^{2,3,4\orcidC{}}$,\newauthor
G. Bruzual$^{5\orcidD{}}$,
S. Charlot$^{6\orcidE{}}$,
G. Ramos-Larios$^{7,8\orcidG{}}$,
L.\,M. Oskinova$^{1\orcidF{}}$,\newauthor  
A.\,A.\,C. Sander$^{9\orcidH{}}$,
S. Reyero Serantes$^{1\orcidI{}}$
\\
$^{1}$Institute for Physics and Astronomy, Universit\"{a}t Potsdam, Karl-Liebknecht-Str. 24/25, D-14476 Potsdam, Germany\\
$^{2}$Instituto Nacional de Astrof{\'\i}sica, \'Optica y Electr\'onica, Luis Enrique Erro 1, Tonantzintla 72840, Puebla, Mexico\\
$^{3}$Consejo Nacional de Humanidades, Ciencias y Tecnolog\'ias, Av. Insurgentes Sur 1582, 03940, Mexico City, Mexico\\
$^{4}$Centro de Estudios de F\'isica del Cosmos de Arag\'on (CEFCA), Plaza San Juan, 1, 44001, Teruel, Spain\\
$^{5}$Instituto de Radioastronom\'{i}a y Astrof\'{i}sica, UNAM Campus Morelia, Apartado postal 3-72, 58090 Morelia, Michoac\'{a}n, Mexico\\
$^{6}$Sorbonne Universit\'e, CNRS, UMR7095, Institut d'Astrophysique de Paris, F-75014, Paris, France \\
$^{7}$Instituto de Astronom\'\i a y Meteorolog\'\i a, CUCEI, Univ.\ de Guadalajara, Av.\ Vallarta 2602, Arcos Vallarta, 44130 Guadalajara, Mexico\\
$^{8}$CUCEI, Universidad de Guadalajara, Blvd. Marcelino Garc\'\i a Barrag\'an 1421, 44430, Guadalajara, Jalisco, Mexico \\
$^{9}$Zentrum f{\"u}r Astronomie der Universit{\"a}t Heidelberg, Astronomisches Rechen-Institut, M{\"o}nchhofstr. 12-14, 69120 Heidelberg, Germany
}

\begin{document}
%\pagerange{\pageref{firstpage}--\pageref{lastpage}}
\maketitle

\begin{abstract}
We present the analysis of archival Very Large Telescope (VLT) Multi-Unit Spectroscopic Explorer (MUSE)
observations of 179~\hii\ regions in the star-forming double-ring collisional galaxy AM\,0644-741 at 98.6~Mpc.
We determined ionic abundances of He, N, O and Fe using the direct method for the brightest \hii\ region (ID\,39);
we report
$\log\rm{(\frac{N}{O})}=-1.3\pm0.2$ and
$12+\log\rm{(\frac{O}{H})}=8.9\pm0.2$.
We also find the so-called `blue-bump', broad \heiiwr, in the spectrum of this knot of massive star-formation;
its luminosity being consistent with the presence of $\sim430$ Wolf-Rayet (WR) stars of the Nitrogen late-type.
We determined the O abundances for 137~\hii\ regions using the strong-line method;
we report a median value of $12+\log\rm{(\frac{O}{H})}=8.5\pm0.8$.
The location of three objects, including the WR complex, coincide with that of an Ultra Luminous X-ray source.
Nebular \heii\ is not detected in any \hii\ region.
We investigate the physical mechanisms responsible for the observed spectral lines using appropriate diagnostic diagrams and ionization models.
We find that the \hii\ regions are being photoionized by star clusters with ages $\sim2.5-20$~Myr and ionization potential $-3.5<$\logU$<-3.0$.
In these diagrams, a binary population is needed to reproduce the observables considered in this work.
\end{abstract}


\section{Introduction}
\label{introduction}

Ring galaxies are a class of peculiar galaxies \citep[][]{Arp1966}
that have apparent ring-like morphologies in optical images \citep[][]{ArpMadore1987}.
Nowadays, ring galaxies are understood as the outcome of a rare-type (1 in 1000) of a head-on
collisional event: a satellite compact galaxy crashes through another's primary disk, and the resulting ring
is the conspicuous short-lived evidence of this collision
\citep[$t_\text{dyn}\sim10^{8}$~yr;][]{Madore2009}.
% Andreas:
Ring galaxies thus are an excellent laboratory to study the
evolution of galaxies, particularly the triggering and suppression
of star formation on galactic scales after such interactions \citep[][]{Higdon2011}.
% Andreas:
In the most recent atlas of collisional ring galaxies,
%In  recent review of this subject,
\citet{Madore2009} report $127$ ring galaxies known so far in the local Universe, of which 108 are located in the southern hemisphere and 19 have positive declinations. % \citep[][]{Arp1966,Appleton1987}.
However, the search for ring galaxies is far from complete and we could expect at least a similar number in the northern hemisphere.
Projects like DESI Legacy Imaging Surveys \citep[][]{DESI2019}
could help with this purpose in the near future.

In order to analyse the star-formation triggered in the rings of these systems
we started a series of studies of ring-type galaxies with available integral field spectroscopy (IFS) observations.
In \citet[][]{Zaragoza-Cardiel+2022} and \citet[][]{Mayya2023} we focused on Cartwheel, the archetypal ring galaxy and the most observed and modelled galaxy of its category \citep[e.g.,][]{Higdon1995,Renaud2018}.
However, there is an even rarer and short lived collisional byproduct subclass:
the "double-ringed" galaxies, which has also been predicted in numerical
simulations by \citet[][]{Lynds1976}.
Here we study the double-ringed galaxy AM\,0644-741, a.k.a the Lindsay-Shapley (L-S) ring, located at a distance of 98.6~Mpc, assuming a Hubble constant H$_{0}=67.8$~\kmsMpc, $\Omega_{m}=0.308$, $\Omega_{\Lambda}=0.692$ (NASA NED\footnote{The NASA/IPAC Extragalactic Database (NED) is funded by the National Aeronautics and Space Administration and operated by the California Institute of Technology.}).

\begin{figure*}
\begin{centering}
\includegraphics[width=0.9\linewidth]{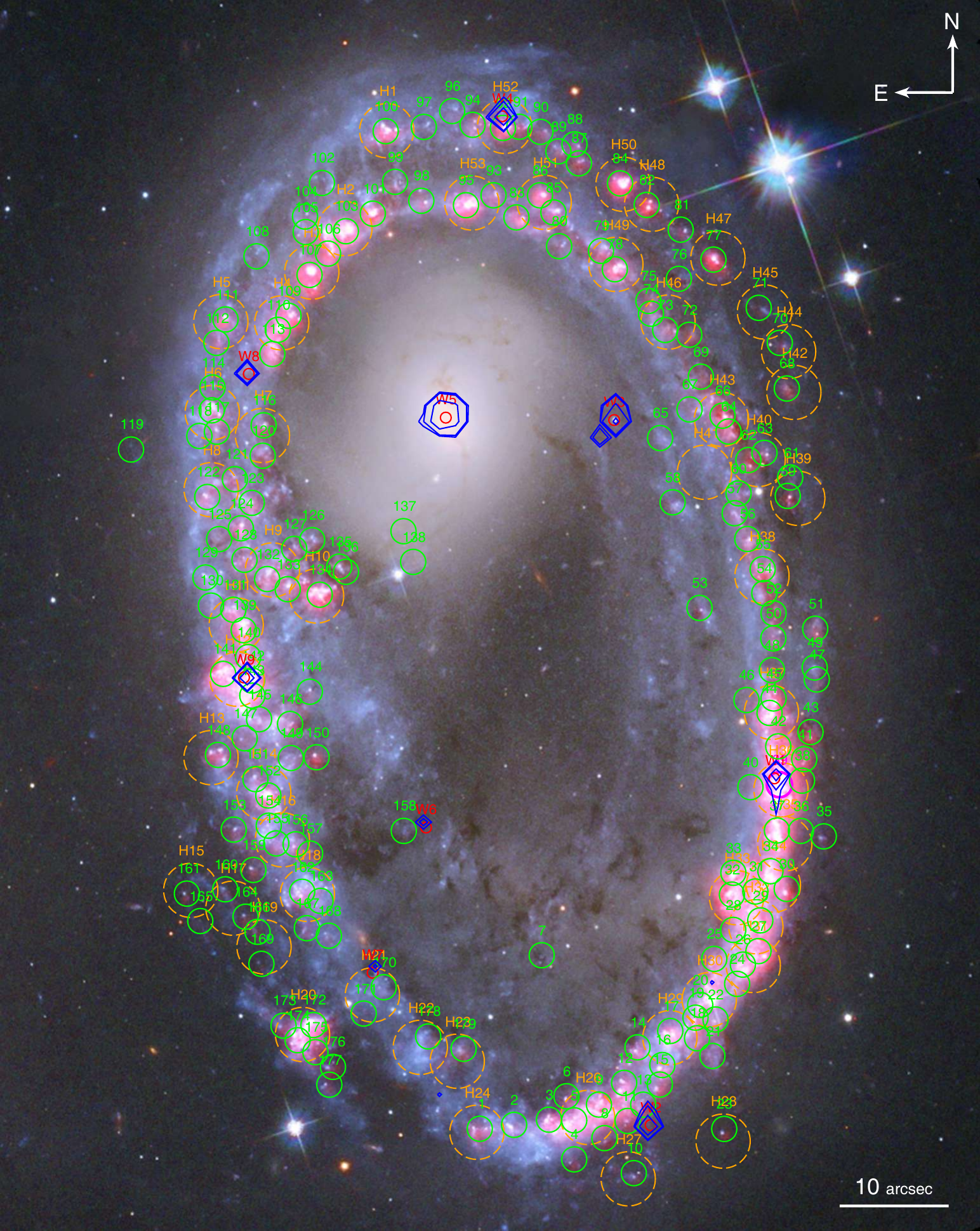}
\par\end{centering}
\caption{
Colour-composite image of the double-ring galaxy AM\,0644-741, formed using the HST/ACS/WFC
images in the filters F814W, F625W and F435W as red, green and blue components,
respectively; and a new \ha\ image constructed from MUSE data (FWHM=1.161~arcsec) as a fourth reddish component.
The 179 \hii\ regions studied here are identified by green circles of 1.162~arcsec radius ($\sim$556~pc at D$=98.62$~Mpc), which is twice the sky resolution size of MUSE.
Star forming regions from \citet[][]{Higdon1997} are indicated in dashed orange circles (H1--H53), and the location of X-ray sources, some of them ULXs, from \citet{Wolter+2018} by red circles (W1--W9) and their
positional error ellipses with blue contours.
The majority of \hii\ regions identified with MUSE are associated with a star-formation complex resolved by HST.
Although ULX sources are located in the star-forming ring, its association with a cluster is not straightforward. The brightest \hii\ region, ID\,39, is also a WR cluster coinciding with the location of the most luminous ULX (W1). See text for details.
}
\label{fig:muse_image}
\end{figure*}

AM\,0644-741 is characterized by its striking double-ring of \hii\ regions and star cluster complexes
noticeable in optical bands, particularly in \ha\ (see Fig.~\ref{fig:muse_image}), but also in the near and far-UV, tracing recent bursts of star formation.
The ring, presumably driven by a radially expanding density wave triggered by the collision of a satellite
object nearly perpendicular to the plane of its disk and close to its centre
\citep[][]{Lynds1976,Higdon2011},
has an angular diameter of 97~arcsec, the second largest (after AM\,1724-622 with 125~arcsec) reported in the Southern Catalogue of ring galaxies by \citet[][]{Madore2009},
corresponding to a physical size of $\sim46.4$~kpc, using the mentioned distance.
A slightly lower value for the major-axis is reported by \citet[][see Tab.~\ref{tab:parameters}]{Higdon2011}, but differences are minor.
The nucleus is north of the geometric center of the elliptical ring,
in the direction of the major-axis.
Compared to the ring, the inner disk is distinctively
more red and gas depleted, with a presumed older stellar population, given the absence of HII regions ionized by young stars.
Unlike its "cousin" the Cartwheel, AM\,0644-741 does not show
a network of spokes in its inner disk, which is practically transparent in the referred bands.
\citet[][]{Few1982} reported that the double-ring is rotating at $311\pm19$~\kms\ and is
expanding at a velocity ($V_\text{exp}$) of $128\pm12$~\kms,
giving a dynamical age ($R_\text{ring}/V_\text{exp}$)$\sim175$~Myr.
In Tab.~\ref{tab:parameters} we list the main physical parameters of AM\,0644-741 collected from literature.

According to the \citet[][]{Madore2009} atlas,
every ring galaxy has a likely collider detected in optical bands.
AM\,0644-741 is not an exception, having at least three identified candidates:
C1, C2 and C3, with a projected separation from the ring nucleus of 78, 110 and 326~arcsec, respectively,
and heliocentric velocities of 7000, 6750 and 6430 \kms, respectively \citep[see Tab.~1 and Fig.~5 in][]{Madore2009}.
Assuming roughly the same distance as the L-S ring,
this is equivalent to a separation between the galaxies of 37, 53 and 156~kpc, respectively.
So far, the true collider has not yet been identified and this issue remains unresolved.

\citet[][]{Higdon1997} were the first to report the massive star formation distribution and
the basic properties of 54 star forming regions in this galaxy; 
we indicate these regions in Fig.~\ref{fig:muse_image} (orange dashed circles) as a reference.
Later, \citet[][]{Higdon2011} studied the impact of star formation on the neutral interstellar medium (ISM).
Since then, no comprehensive study has been carried out on this galaxy.
The availability of spectrographs incorporating integral field units (IFUs) on
large telescopes such as Multi Unit Spectroscopic Explorer (MUSE) on the 
Very Large Telescope (VLT) \citep{Bacon+2010} allows a new in-depth study of AM\,0644-741.
One MUSE dataset is available on this galaxy at the
seeing-limited spatial resolution of 1.161~arcsec. This dataset provides 
optical spectra covering a range of 4750 to 9350~\AA\
(corresponding to a rest wavelength range of 4650 to 9250~\AA, considering its redshift: z=0.022)
over the entire galaxy.

\begin{table}
\small\addtolength{\tabcolsep}{-0.5pt}
  \caption{Parameters of the double-ringed galaxy AM\,0644-741.}
\setlength{\tabcolsep}{0.2\tabcolsep}
  \label{tab:parameters}
\begin{tabular}{ccc}
\hline
Parameter & Value & Ref. \\
\hline
Cross-identifications                     & ESO\,034-IG 011      & (a) \\
                                          & IRAS\,06443-7411     & (a) \\
Galaxy type                               & Lindsay-Shapley ring & (a) \\
Right ascension (J2000)                   & 06h43m06.1s          & (a) \\
Declination (J2000)                       & $-$74d13m35s         & (a) \\
z (Helio)                                 & $0.02203\pm0.00009$  & (b) \\
Velocity (Heliocentric) [km/s]            & $6604\pm26$          & (b) \\
Hubble distance (CMB) [Mpc]               & $98.62\pm6.91$       & (a) \\
Foreground Gal. ext. (\Av) [mag]          & 0.3537               & (c) \\
Z [Z$_{\odot}$] (Bulge spectra)           & 0.45                 & (d) \\
Angular diameter [arcsec]                 & $95\times52$         & (e) \\
Physical diam. [kpc]                      & $45\times25$         &  \\
L$_{\text{\ha}}$ [$10^{42}$ erg s$^{-1}$] & $1.40\pm0.06$        & (e) \\
L$_{\text{X}}$ (0.5-10 keV) [$10^{41}$ erg s$^{-1}$] & 1.78      & (g) \\
EW$_{\text{\ha}}$ [\AA]                   & 14                   & (f) \\
SFR [M$_{\odot}$~yr$^{-1}$] (from L(\ha)) & $11.2\pm0.4$         & (e) \\
SFR [M$_{\odot}$~yr$^{-1}$] (from radio)  & $17.6\pm0.9$         & (g) \\
$V_\text{rot}$ [\kms]                     & $311\pm19$           & (h) \\
$V_\text{exp}$ [\kms]                     & $128\pm12$           & (h) \\
\hline
\end{tabular}

(a) NASA NED;
(b) \citet[][]{Fisher1995};
(c) \citet[][]{SchlaflyFinkbeiner2011};
(d) \citet[][and references therein]{Mapelli+2009};
(e) \citet[][]{Higdon2011};
(f) \citet[][and references therein]{Higdon1997};
(g) \citet[][]{Wolter+2018};
(h) \citet[][]{Few1982}.
\end{table}

%Divakara:
MUSE datacubes of ring galaxies offers the opportunity to address several problems that are particular to the ring galaxies.
The first of those is the study of metallic abundances and its radial gradient.
Also, ring galaxies are known to be prolific sources of Ultra Luminous X-ray (ULX) sources \citep[][]{Wolter+2018}.
MUSE data offers an opportunity to study the impact of these sources on the state of high ionization potential emission lines from \ariv\ and \heii.
X-ray sources are often postulated to be responsible of these ions \citep[][]{Schaerer2019}.
Furthermore, star-forming regions pass though the WR phase for a short term during which time the blue part of the spectrum is expected to contain the so-called `blue-bump' (hereafter, BB), made up basically of broad emission lines such as \heii\ and, depending on the WR-type, with contributions of \ciii-\civ\ or \niii-\nv. However, the detection of BB is heavily metallicity dependent.
All these phenomena have been addressed in the Cartwheel using the MUSE  data in our previous papers \citep[][]{Mayya2023,Zaragoza-Cardiel+2022}. For example, in the Cartwheel, which has around 1/4 \zsol,
we found enrichment of O in the ring \hii\ regions without the corresponding enrichment
in N and Fe, metals expelled by intermediate-age and low-mass stars.
In the Cartwheel, we did not find signature of X-ray ionization even in \hii\ regions coinciding with ULXs, and those where \heiiwr\ line is detected.

We employ the available MUSE observations to continue our series of studies on the star-formation triggered in the ring of this type of galaxies.
Here in particular, we study the ionizing mechanisms and the chemical abundances
in the star cluster complexes hosted in the \hii\ regions that shape the double-ringed morphology of AM\,0644-741.
We are interested in comparing our results in this galaxy with those obtained in the Cartwheel,
to better understand the nature and evolution of ring structures in galaxies
with different environments, star formation histories and metallicities in its post-collisional stage.
Thanks to new spectroscopic data cubes like those of MUSE/VLT these studies are now possible.
In a \ha\ image we constructed using the MUSE dataset (see Sec.\,\ref{Spectroscopic data})
we identified 179 \hii\ regions in the double-ring of AM\,0644-741.
At the distance of the L-S ring, MUSE spectra are available at physical 
scales of $\sim$556~pc.
At the spatial resolution of the Advanced Camera for Surveys (ACS) Wide Field Channel (WFC)
images of the Hubble Space Telescope (HST), which are the highest resolution images
available for AM\,0644-741 (0.05~arcsec/pix), we can associate 
each MUSE-identified \hii\ region with a population of super star clusters, which presumably provide the ionization of the \hii\ regions.
A colour-composite image using our MUSE \ha\ image and HST filters is shown in Fig.~\ref{fig:muse_image}.
All images used in this figure are astrometrized as explained in Section 2.2. However, the point spread function (PSF) of the images are not matched and correspond to the respective observed values given in the section below.

This article is structured as follows:
in Sec.~\ref{Spectroscopic data} we describe the spectroscopic dataset, the constructed catalog of \hii\ regions, the extraction of the spectra
and measurement of the emission line fluxes;
in Sec.~\ref{Determination of chemical abundances} we determine the abundances of our objects;
we study the ionization sources with diagnostic diagrams and models in Sec.~\ref{Ionizing sources};
our results are discussed in Sec.~\ref{Discussion}; 
finally, a summary and our conclusions are given in Sec.~\ref{Summary and concluding remarks}.

\section{Spectroscopic data}\label{Spectroscopic data}
\subsection{VLT/MUSE observations}\label{VLT/MUSE observations}

\begin{figure*}
\begin{center}
\includegraphics[width=1\linewidth]{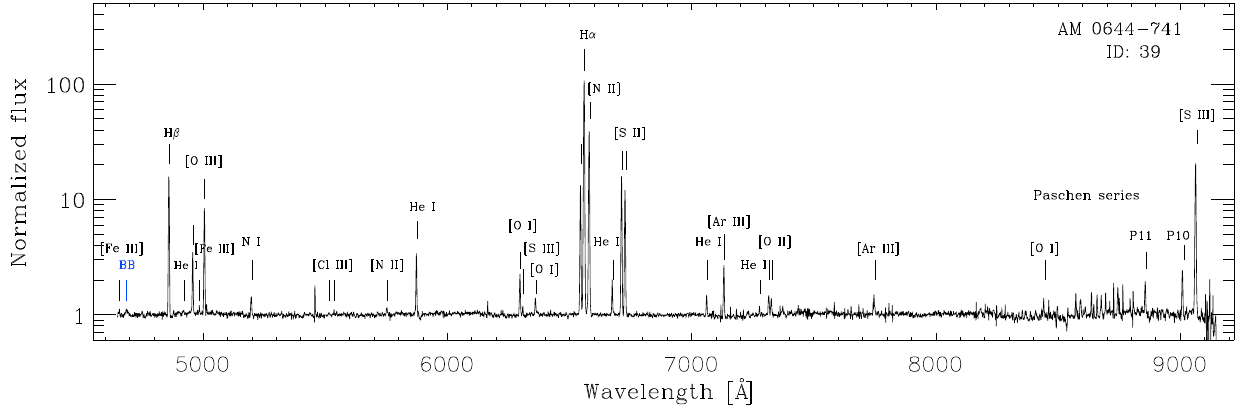}
\caption{VLT MUSE redshift-corrected spectrum of the cluster complex ID\,39, WR1, located in the SW section of the ring of the AM\,0644-741 galaxy (see Fig.~\ref{fig:muse_image}).
The so-called `blue-bump' (BB) WR feature at 4686~\AA, as well as the most common nebular lines of typical \hii\ regions,
from \feiii\ at 4658~\AA\ to \siii\ at 9068.6~\AA, are indicated.
The spectrum is shown normalized to its best-fit continuum.
}
\label{fig:MUSE_spec}
\end{center}
\end{figure*}

MUSE is a panoramic integral-field spectrograph at the 8-m VLT of the European Southern Observatory (ESO)\footnote{\url{http://muse-vlt.eu/science/}}, operating in the optical wavelength range from $\sim$4750 to $\sim$9351~\AA\, with a spatial sampling of 0.2 arcsec\,pixel$^{-1}$, a spectral sampling of 1.25~\AA\,pixel$^{-1}$ and a spectral resolution FWHM$\sim$3~\AA\ \citep{Bacon+2010}.
This spectral resolution corresponds to $\sigma\sim$40--60~\kms, which is not good enough to resolve the lines but sufficient to detect the strong lines to measure abundances.
The public MUSE data cube of AM\,0644-741 galaxy was retrieved from the ESO Science Archive Facility\footnote{\url{http://archive.eso.org/cms.html}}.
The data cube was already flux and wavelength calibrated, ready for scientific exploitation.
The data were obtained during a single observing run on November 9, 2020;
the  Principal Investigator (PI) is Bernd Husemann (Program ID: 106.2155);
dedicating a total exposure time of $=$1350~s (effective time of $=$714.5~s) in a single observation.
A sky coverage of observations of 2.1~arcmin$^{2}$,
encloses the entire star-forming ring of the AM\,0644-741 galaxy,
as can be seen in Fig.~\ref{fig:muse_image}; here we show an RGB image formed by combining
the HST/ACS/WFC filters in F814W, F625W and F435W as red, green and blue components, respectively;
the \ha\ image constructed from MUSE data is used as a fourth reddish component.
The MUSE images have a seeing-limited resolution of Full Width at Half Maximum (FWHM) of 1.162~arcsec, which corresponds to a spatial scale of $\sim$556~pc at the distance of the L-S ring galaxy. We list the main observation details in Tab.~\ref{tab:muse}.

\subsection{MUSE data cube and astrometry}\label{MUSE data cube and astrometry}

We used the reduced 3-D data cube as a starting point in our analysis.
The tool QFitsView\footnote{\url{https://www.mpe.mpg.de/~ott/dpuser/qfitsview.html}}
\citep{Ott2012} was used for the extraction of images at selected wavelengths,
as well as spectra of selected regions.
A continuum image covering the entire spectral range of MUSE is prepared to transform MUSE images into International Celestial Reference System (ICRS) as defined by the coordinates of the point sources in the 2MASS catalogue. The Field of View (FoV) of the MUSE image of AM\,0644-741 contains three 2MASS stars which is insufficient to obtain accurate solutions.
We hence carried out the astrometry in two steps, first we brought the HST F814W images into ICRS using 14 2MASS stars, and then we identified 13 compact objects in the MUSE continuum image and the F814W image, which we used to find astrometric solution of the MUSE image.
The astrometric accuracy of the resulting image is 0.1~arcsec. We used the {\scshape iraf} tasks {\it ccmap} and {\it wregister} to carry out the astrometric calibration.

\begin{table}
\small\addtolength{\tabcolsep}{-0.5pt}
\begin{center}
\caption{Details of the VLT-MUSE dataset${^\dagger}$ of AM\,0644-741.}
\begin{tabular}{ll}
\hline
Parameter               &                \\
\hline
R.A. (J2000)              & 06:43:05.01    \\
Dec. (J2000)             & --74:14:28.4   \\
Date of observation     & 09/11/2020     \\
Sky coverage            & 2.1~arcmin$^{2}$    \\
Sky resolution$^{\dagger\dagger}$          & 1.162~arcsec   \\
Pixel scale             & 0.2~arcsec     \\
Spectral range          & 4750-9350~\AA\ \\
Spectral resolution (R) &  3027         \\
Effective exposure time & 714.5~s        \\
Sensitivity (AB mag at 5$\sigma$) & 20.87~mag \\
\hline
\end{tabular}\\
${^\dagger}$ Single IFU cube; data processing certified by ESO;
PI: B. Husemann;
OB ID: 2885683.
$^{\dagger\dagger}$ FWHM effective spatial resolution.
\label{tab:muse}
\end{center}
\end{table}

\subsection{Catalog}\label{Catalog}

For our study in AM\,0644-741 it was necessary to build a new catalog of \hii\
regions using the available MUSE observations.
As a first step, we generated a continuum-free \ha\ image from the MUSE datacube using QFitsView. For achieving this, we summed fluxes in a 8.8~\AA\ window centered on \ha\ at the redshifted wavelength of this galaxy subtracting a continuum 44~\AA\ away on either side of the \ha\ line. With these values we ensure that the emission line is covered by the width of the filter, taking into account the shift of the lines due to its kinematics.

The continuum-subtracted \ha\ image (the reddish component in Fig.~\ref{fig:muse_image}) allows the identification of \hii\ regions using source extractor codes such as {\sc pyhiiextractor} \citep{Lugo2022}, SExtractor \citep{Bertin1996}, etc. The former code measures accurate fluxes of \hii\ regions, taking into account corrugated boundaries. We opted to measure fluxes by fitting a Gaussian function to line profiles on spectra extracted using circular apertures of fixed radius around the identified regions. This method ensures that the fluxes of all lines are measured over the same areas. For this reason, we used {\sc SExtractor}
to identify the locations
of the objects for later extraction of their spectra.
We identified a total of 184 regions with \ha\ emission in the entire FoV of this galaxy.
Then, spectra were extracted with apertures of 1.162~arcsec radius, which is equivalent to 1 seeing FWHM, for all regions automatically selected.
Finally, after careful inspection of the spectra, five objects were later identified to be spurious sources (which include the nucleus of the galaxy and probable background stars), ultimately resulting in a catalog of 179 \hii\ regions.
Most of these sources are located in the double-ring of the galaxy (see Fig.~\ref{fig:muse_image}).
The spectrum of the most conspicuous \hii\ region in AM\,0644-741 is shown in Fig.~\ref{fig:MUSE_spec}.
The foreground Galactic extinction in the direction of AM\,0644-741 is $\text{A}_{\text{V}}=0.35$~mag
\citep[][]{SchlaflyFinkbeiner2011}.
The extracted spectra were first corrected for this foreground extinction using the Galactic reddening curve  of \citet[][]{Cardelli+1989}. The selected regions are at least an order of magnitude brighter in surface brightness with respect to the diffuse ionized gas (DIG), implying that the contamination of the observed line ratios by the DIG is negligible. Moreover with the recent understanding that the DIG in star-forming galaxies is created due to photoionization by the photons leaking the \hii\ regions \citep{Della2020, 2022Belfiore}, it is expected to occupy the immediate surroundings of \hii\ regions in the disk rather than the regions along the line of sight to \hii\ regions. Hence, we did not apply any correction to the observed line ratios.

AM\,0644-741 has nine identified X-ray sources of which, excluding the galaxy nucleus and a background 
active galactic nucleus (AGN),
seven were identified as ULXs by \citet{Wolter+2018}.
The ULX sources are located in the ring of the galaxy
and some of them coincide with the projected location of an \hii\ region hosting a star-forming complex (see Fig.~\ref{fig:muse_image}).
At least three of the ULXs in the ring are associated with a \hii\ region of our sample.
We address the nature and probable role of the ULXs in the ionization of their host \hii\ regions in Sec.\,\ref{Ionizing sources}.

\begin{table}
\small\addtolength{\tabcolsep}{-0.5pt}
\begin{center}
\caption[]{
De-reddened nebular line fluxes relative to \hb=100 for the \hii\ region ID\,39 in AM\,0644-741 (see Fig.~\ref{fig:muse_image}). Its spectrum is shown in Fig.~\ref{fig:MUSE_spec}. The complete information for all the \hii\ regions is given as extra material.
}
\setlength{\tabcolsep}{0.4\tabcolsep}
\begin{tabular}{ccc}
\hline
$\lambda_{0}$    &Ion& Flux             \\ 
(\AA)  &         &  (2.3 arcsec)    \\ 
\hline
4658.0 & \feiii\ & 0.55$\pm$0.17  \\  
4711.4 & \ariv\  & 0.48$\pm$0.15           \\ 
4740.2 & \ariv\  & 0.47$\pm$0.15              \\ 
4861.4 & \hb\    & 100                \\ 
4921.9 & \hei\   & 0.38$\pm$0.13   \\  
4958.9 & \oiii\  & 14.87$\pm$0.16  \\ 
4985.9 & \feiii\ & 1.02$\pm$0.18   \\ 
5006.8 & \oiii\  & 43.71$\pm$0.04    \\   
5200.0 & \nia\   & 2.92$\pm$0.21     \\  
5517.7 & \cliii\   & 0.20$\pm$0.08   \\  
5754.6 & \nii\   & 0.72$\pm$0.17   \\  
5875.6 & \hei\   & 8.58$\pm$0.22   \\ 
6300.0 & \oi\    & 3.82$\pm$0.15  \\ 
6312.1 & \siii\  & 0.45$\pm$0.14            \\ 
6363.8 & \oi\    & 1.31$\pm$0.15   \\     
6548.1 & \nii\   & 33.35$\pm$0.04   \\   
6562.8 & \ha\    & 287$\pm$2.25     \\   
6583.5 & \nii\   & 101.91$\pm$0.69    \\  
6678.2 & \hei\   & 2.34$\pm$0.13    \\ 
6716.4 & \sii\   & 38.41$\pm$0.07   \\  
6730.8 & \sii\   & 28.10$\pm$0.05     \\   
7065.2 & \hei\   & 1.07$\pm$0.19     \\ 
7135.8 & \ariii\ & 3.75$\pm$0.13     \\ 
7170.5 & \ariii\ & 0.16$\pm$0.09     \\ 
7281.3 & \hei\   & 0.32$\pm$0.09       \\ 
7319.2 & \oii\   & 1.26$\pm$0.15     \\ 
7330.2 & \oii\   & 0.97$\pm$0.16     \\ 
7751.1 & \ariii\ & 1.17$\pm$0.17      \\ 
8446.4 & \oi\    & 0.42$\pm$0.22      \\ 
8665.0 & P\,13   & 0.46$\pm$0.17   \\ 
8750.5 & P\,12   & 1.02$\pm$0.18  \\ 
8862.9 & P\,11   & 1.22$\pm$0.22   \\ 
9015.3 & P\,10   & 1.59$\pm$0.19      \\ 
9068.6 & \siii\  & 19.57$\pm$0.36     \\ 
\hline
$\log\rm{(F(H}\beta$)) & & $-13.567\pm0.007$ erg~cm$^{-2}$~s$^{-1}$ \\ 
$A_{\rm V}$      & & 1.225$\pm$0.014~mag  \\  
$c$(H$\beta$)    & & 0.730$\pm$0.007~mag    \\  
EW(H$\beta$)     & & 55.75~\AA\          \\ 
SNR(H$\beta$)    & & 141.2               \\
\hline
\end{tabular}
\vspace{0.4cm}
\label{tab:lines}
\end{center}
\end{table}

\subsection{Measurement of line fluxes}\label{Measurement of line fluxes}

The analysis of the ionizing mechanisms of the \hii\ regions in the double-ring
of AM\,0644-741 and the determination of their nebular abundances requires the
previous determination of de-reddened flux ratios of the nebular lines relative
to the flux of \hb.
Considering that our sample of \hii\ regions might be affected by the presence of
an underlying absorption feature (e.g. Balmer lines, \hb\ in particular),
first we fitted their continuum with Single Stellar Population (SSP) synthesis models.
We used the GIST pipeline \citep{2019A&A...628A.117B},
making use of the most recent Charlot \& Bruzual SSP models (Charlot \& Bruzual, in preparation), to fit the continuum and the absorption lines over the entire observed wavelength range of our spectra: 4650--9250~\AA, where the MUSE spectral resolution is $\sim3$~\AA.
Then we subtracted the fitted continuum to obtain pure nebular spectra,
which is the one we used for later analysis.

Around 30 nebular lines were identified, from \feiii\ at 4658~\AA\ to \siii\ at 9068.6~\AA, in the spectrum of the brightest \hii\ region in AM\,0644-741, ID\,39,
most of which have been marked in Fig.~\ref{fig:MUSE_spec}.
These lines were analysed using the Gaussian line fitting option of the {\scshape iraf} task {\it splot}.
The measured line fluxes were corrected for
extinction by using the obtained $c$(H$\beta$) value assuming an intrinsic Balmer-decrement ratio of H$\alpha$/H$\beta=2.86$, for a case B photoionised nebula,
corresponding to mean values for $T_{\rm e}$ in the range 2000~K and 12000~K
and $n_{\rm e}$ between 100~cm$^{-3}$ and 1200~cm$^{-3}$ \citep[see][]{OsterbrockFerland2006}.
We used the reddening curve of \citet{Cardelli+1989} with $R_\mathrm{V}$=3.1.
The resultant line fluxes and statistical errors are listed in Tab.~\ref{tab:lines} for the brightest \hii\ region of the sample.

The spectrum of ID\,39 shows a Wolf-Rayet (WR) feature, the BB, at 4686~\AA\ \citep[][]{Allen+1976,KunthSargent1981}.
This feature has been observed both in metal-rich \citep[e.g. the Antennae galaxies by][at~18.1 Mpc]{Gomez-Gonzalez+2021}, LMC-like \citep[e.g. NGC\,3125 by][at 11.5~Mpc]{Hadfield2006} and metal-poor galaxies \citep[e.g. Mrk\,178, NGC\,625 by][at 3.9~Mpc]{Kehrig2013, Monreal2017}.
Recent BB detections also include M\,81 \citep[][at 3.6~Mpc]{Gomez-Gonzalez+2016}
We address the nature of the BB in ID\,39 (hereafter WR1) in Sec.\,\ref{Wolf-Rayet stars}.

\begin{figure}
\begin{centering}
\includegraphics[width=1\linewidth]{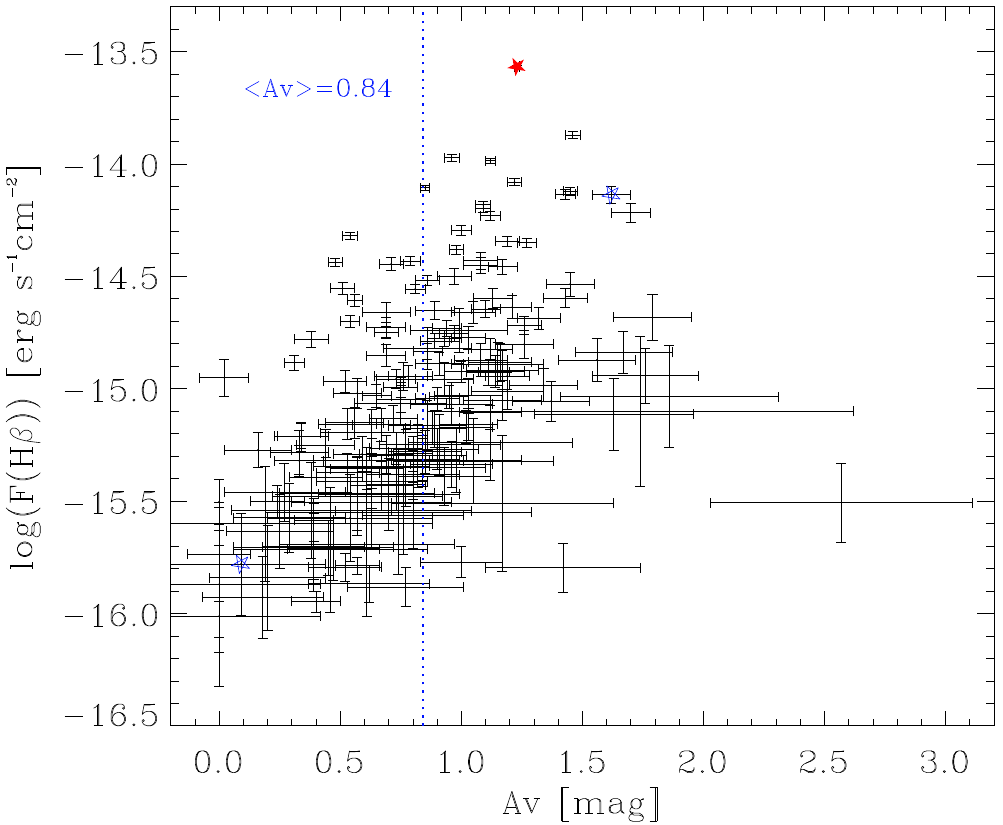}~
\par\end{centering}
\caption{Distribution of the visual extinction \Av\ and the reddening
corrected F(\hb) of the \hii\ regions in the double-ring of AM\,0644-741.
The average value of \Av$=0.84$~mag is shown by the vertical dashed line.
\Av\ tends to increase with F(\hb).
The star symbols indicates the \hii\ regions with a ULX counterpart (ID\,39, 92 and 142).
The red star represents the \hii\ region (ID\,39) with a WR feature in its spectrum.
}
\label{fig:av}
\end{figure}

Reddening-corrected fluxes for all the objects are reported in a Table as extra-material.
The distribution of the visual extinction (\Av) and the reddening corrected F(\hb) is shown in Fig.~\ref{fig:av}.
It can be noticed that
\Av\ tends to increase with F(\hb). A median value of \Av\ $\sim0.84$~mag is indicated by a dotted vertical line.
The correlation between \Av\ and F(\hb) arises because the analyzed \hii\ regions adhere to a Kennicutt–Schmidt star-formation relation \citep{Ballesteros2020}. In other words, the gas density of a star-forming region, which dictates \Av\ through the almost constant gas-to-dust ratio, correlates with the star formation rate (SFR), which is traced by \hb. It can be recalled that Cartwheel \hii\ regions do not show such a correlation \citep{Zaragoza-Cardiel+2022}. The median values obtained here are lower than the corresponding values for the bright ring \hii\ regions in the Cartwheel but are comparable to the median \Av\ of the fainter regions in the outer ring of the Cartwheel.

\section{Determination of chemical abundances}\label{Determination of chemical abundances}
\subsection{Direct Method}\label{Direct method}
Temperature-sensitive lines are detected in only one (ID\,39,) of the 179 \hii\ regions in AM\,0644-741,
making it possible to determine abundance using the direct method (DM) only for this region.
This happens to be the brightest nebular region with
log(F(H$\beta$)/[erg~cm$^{-2}$~s$^{-1}$])$=-13.567\pm0.007$
(see Fig.~\ref{fig:av}) and SNR(\hb)$=141$ (see Tab.~\ref{tab:results})
and host the only regions where a WR feature is detected (see Fig.~\ref{fig:MUSE_spec})
the location of this brightest \hii\ region also coincides with
the brightest ULX (W1) reported in this galaxy (see Fig.~\ref{fig:muse_image}).
Determining the O abundance of this region is key to estimate the WR population responsible for the broad \heiiwr\ line. We address this issue in Sec.\,\ref{Wolf-Rayet stars}.

We followed the methodology used to determine O abundances in the Cartwheel \hii\ regions \citep{Zaragoza-Cardiel+2022}.
First we determined the physical conditions of the ionized gas:
electron temperature ($T_\mathrm{e}$) and electron density ($n_\mathrm{e}$).
$T_\mathrm{e}$ and $n_\mathrm{e}$ are estimated simultaneously with PyNeb \citep{2015A&A...573A..42L}.
We use the three ionization zones approximation from \citet{Garnett1992}:
$T_\mathrm{e}$ for the low ionization zone ($T_{\rm{e}}^{\rm{low}}$)
was determined with \nii($\lambda6548$+$\lambda6584$)/$\lambda5755$;
the medium ionization zone ($T_{\rm{e}}^{\rm{medium}}$) with \siii$\lambda6312$/$\lambda9069$
and the high ionization zone ($T_{\rm{e}}^{\rm{high}}=T_{\rm{e}}^{\rm{[OIII]}}$) using its relation with $T_{\rm{e}}^{\rm{medium}}$  from  \citet{2006MNRAS.372..293H}: 
\begin{equation}
   T_{\rm{e}}^{\rm{medium}}=1.19\thinspace T_{\rm{e}}^{\rm{high}}-3200\rm{K}. 
\end{equation}

Densities determined from \sii$\lambda6717$/$\lambda6731$ were used for all the three zone lines.
In Tab.~\ref{tab:dm} we list our results for $T_\mathrm{e}$, $n_\mathrm{e}$, the ionic abundances of He, N, O and Fe for WR1.

\begin{figure*}
\begin{centering}
\includegraphics[width=0.85\linewidth]{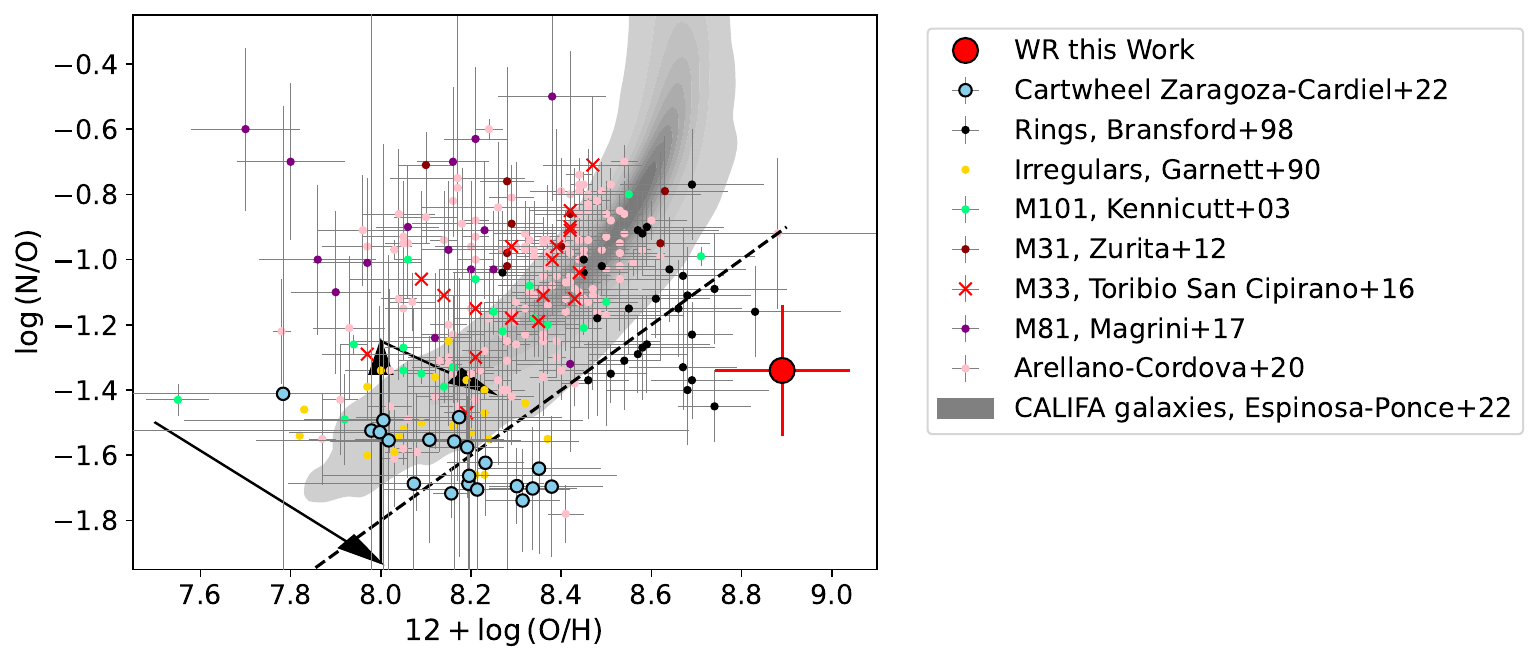}~
\par\end{centering}
\caption{$\log\rm{(\frac{N}{O})}$ vs. $12+\log\rm{(\frac{O}{H})}$ for the WR \hii\ region (red circle) detected in AM\,0644-741.
N and O abundances were determined with the DM.
Following \citet[][]{Zaragoza-Cardiel+2022} we compare our results with the corresponding values reported for \hii\ regions in other systems: Cartwheel \citep[][blue dots]{Zaragoza-Cardiel+2022}; ring galaxies \citep[][black dots]{Bransford1998}; irregular galaxies \citep[][yellow dots]{Garnett1990};
M\,101 \citep[][green dots]{Kennicutt2003};
M\,31 \citep[][red dots]{Zurita2012};
M\,33 \citep[][red crosses]{Toribio2016} and M\,81 \citep[][purple dots]{Magrini2017};
\hii\ regions of several galaxies studied by \citet[][pink dots]{Arellano2020}
and results from the CALIFA survey \citep[][grey region]{Espinosa2020}.
Following \citet[][]{Zaragoza-Cardiel+2022} we indicate the schematic representation of the evolution of the N/O ratio for a dwarf galaxy from \citet[][]{Garnett1990}.
}
\label{fig:abun}
\end{figure*}

\begin{figure*}
\begin{centering}
\includegraphics[width=0.85\linewidth]{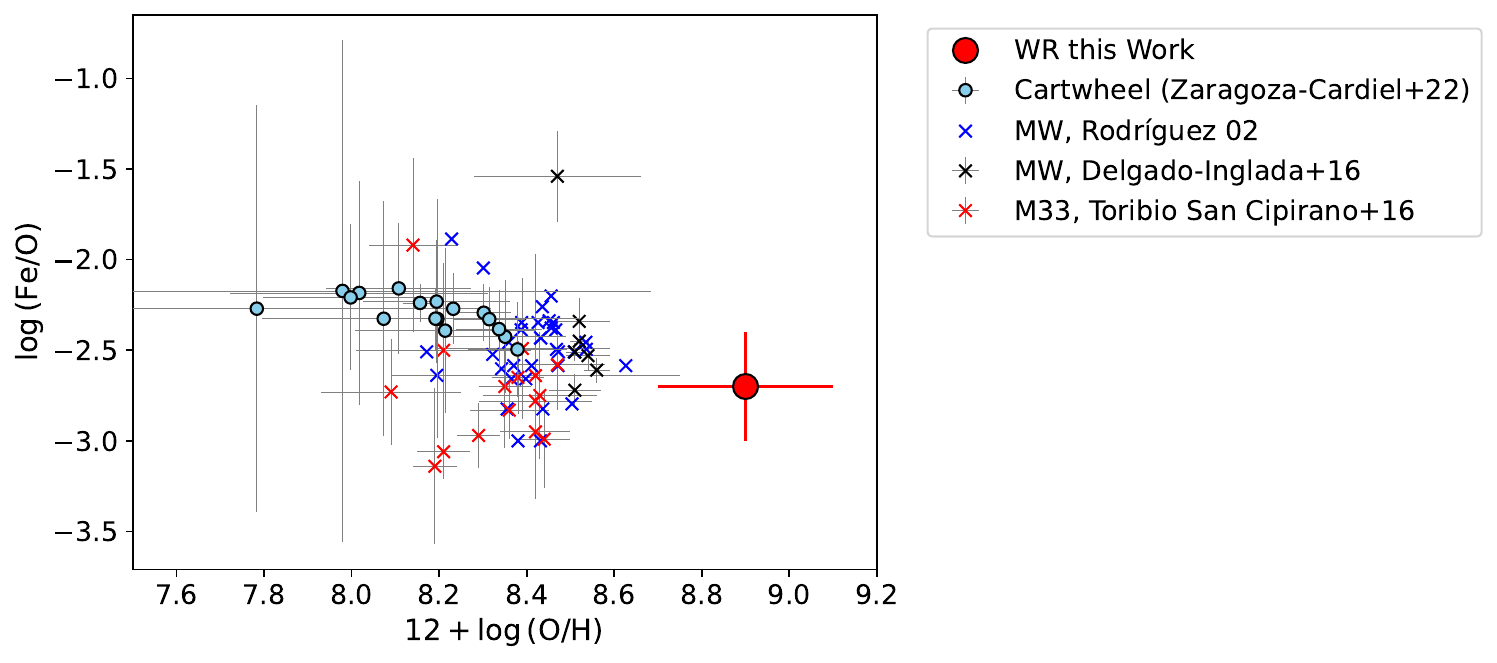}~
\par\end{centering}
\caption{$\log\rm{(\frac{Fe}{O})}$ vs. $12+\log\rm{(\frac{O}{H})}$ for the WR \hii\ region (red circle) detected in AM\,0644-741.
Fe and O abundances were determined with the DM.
Following \citet[][]{Zaragoza-Cardiel+2022} we compare our results with the corresponding values reported for \hii\ regions in other systems: Cartwheel \citep[][blue dots]{Zaragoza-Cardiel+2022};
the Milky Way by \citep[][blue crosses]{Rodriguez2002} and \citet[][red crosses]{Delgado2016} and in 
M\,33 \citep[][red crosses]{Toribio2016}.
}
\label{fig:abunFe}
\end{figure*}

In Fig.~\ref{fig:abun} we compare $\log\rm{(\frac{N}{O})}$ vs. $12+\log\rm{(\frac{O}{H})}$ for the WR \hii\ region
detected in AM\,0644-741 with \hii\ regions in other galaxies:
the prototype ring-galaxy Cartwheel \citep[][]{Zaragoza-Cardiel+2022} and other ring galaxies \citep[][]{Bransford1998}; with irregular galaxies \citep[][]{Garnett1990};
and other well known galaxies like M\,101 \citep[][]{Kennicutt2003},
M\,31 \citep[][]{Zurita2012}, M\,33 \citep[][]{Toribio2016} and M\,81 \citep[][]{Magrini2017},
with \hii\ regions of a sample of galaxies studied by \citet[][]{Arellano2020} and objects from the CALIFA survey \citep[][]{Espinosa2020}. 
Note that all abundances, except those from \citep{Bransford1998} and \citep{Espinosa2020}, are obtained using the DM.
Following \citet[][]{Zaragoza-Cardiel+2022} we also indicate the schematic representation of the expected evolution of the N/O ratio for a dwarf galaxy from \citet[][]{Garnett1990} as a reference.
The N/O ratio of WR1 is close to \hii\ regions in other irregular galaxies. However, the 
O abundance for AM\,0644-741, with $12+\log\rm{(\frac{O}{H})}=8.9$, is higher by as much as 0.6~dex, and is among the highest values measured using the DM. It is likely that the equation that we have used to determine temperature of the high ionization zone, obtained for low metallicity irregular galaxies, underestimates the temperature, thus slightly overestimating the abundances. Temperature calibrations for high metallicity regions are lacking \citep[see e.g.][]{Arellano2020}.

In Fig.~\ref{fig:abunFe} we compare $\log\rm{(\frac{Fe}{O})}$ vs. $12+\log\rm{(\frac{O}{H})}$ for WR1 with the \hii\ regions in Cartwheel \citep[][]{Zaragoza-Cardiel+2022}, in the Milky Way \citep[MW;][]{Rodriguez2002,Delgado2016} and in M33 \citep[][]{Toribio2016}.
WR1 in  AM\,0644-741 follows the trend of the \hii\ regions in these galaxies, the higher the abundance of O, the lower that of Fe. 
However, the measured O abundance in AM\,0644-741 is a clear outlier, which as mentioned above, is likely due to the lack of reliable metallicity calibrations at the high-metallicity end.

\begin{table}
\begin{center}
\caption[]{
$T_{\rm{e}}$, $n_\mathrm{e}$ and He, O, N and Fe ionic abundances using the DM for ID\,39 (WR1)
in AM\,0644-741.
}
\setlength{\tabcolsep}{0.4\tabcolsep}
\begin{tabular}{llc}
\hline
Parameter                         & note                                 & value          \\ 
\hline
$T_{\rm{e}}^{\rm{low}}$\nii\      &                                      & $7726\pm488$~K \\ 
$T_{\rm{e}}^{\rm{medium}}$\siii\  &                                      & $6853\pm641$~K \\
$T_{\rm{e}}^{\rm{high}}$          & relation from \citet{Garnett1992}    & $8448\pm772$~K \\
$n_\mathrm{e}$                    &    $T_{\rm{e}}^{\rm{low}}$\nii\      & $54\pm2$~cm$^{-3}$ \\
$n_\mathrm{e}$                    &    $T_{\rm{e}}^{\rm{medium}}$\siii\  & $53\pm2$~cm$^{-3}$ \\
$12+\log\rm{(\frac{He}{H})}$      & He$^+$ component                     & $10.75\pm0.14$ \\ 
$\log\rm{(\frac{N}{O})}$          &                                      & $-1.3\pm0.2$   \\
$12+\log\rm{(\frac{O}{H})}$       &                                      & $8.9\pm0.2$    \\
$\log\rm{(\frac{Fe}{O})}$         &                                      & $-2.7\pm0.3$   \\ 
\hline
\end{tabular}
\vspace{0.4cm}
\label{tab:dm}
\end{center}
\end{table}

\subsection{Strong-line Method}\label{Strong-line method}

\begin{figure}
\begin{centering}
\includegraphics[width=0.95\linewidth]{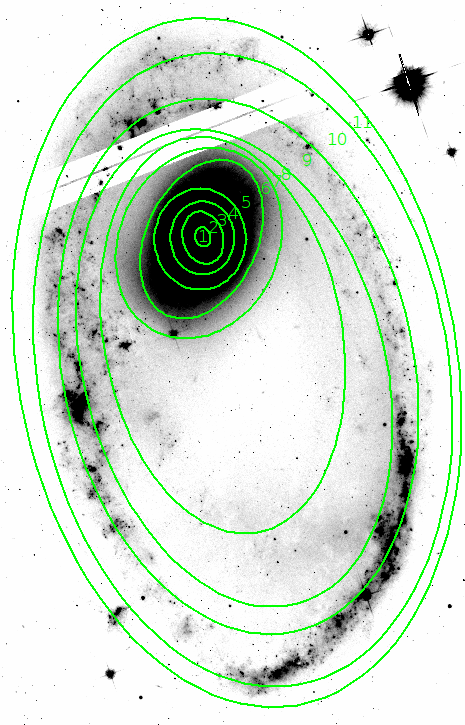}~
\par\end{centering}
\caption{
HST/ACS/WFC image in F625W filter of AM\,0644-741 with the ellipses ($1-11$) used to estimate the galactocentric radius and position angle of the \hii\ regions in the star-forming double-ring of this galaxy.
}
\label{fig:elipses}
\end{figure}

\begin{figure}
\begin{centering}
\includegraphics[width=1\linewidth]{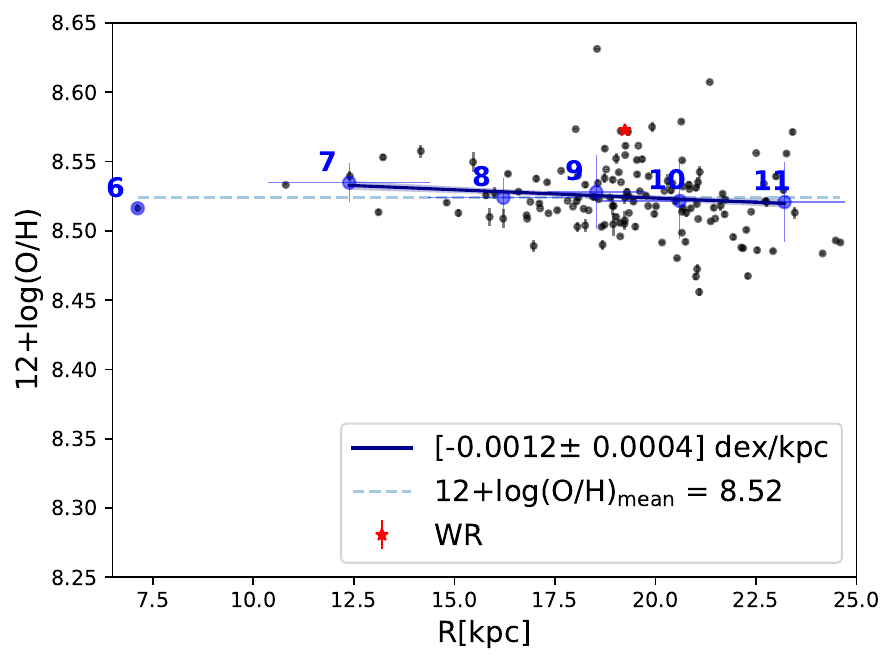}~
\par\end{centering}
\caption{$12+\log\rm{(\frac{O}{H})}$ vs. galactocentric radius (R) for the \hii\ regions (black points) located in the double ring of AM\,0644-741.
Average values for the regions amidst the ellipses defined in Fig.~\ref{fig:elipses} are indicated with blue points.
O abundances were determined with the SLM;
the blue dashed line indicates the median value.
A linear fit (blue line) to the L-S ring data points indicates a slight gradient of
$-0.0012\pm0.0004$~dex/kpc among the ellipses 7 and 11.
The red star represents the \hii\ region  (ID\,39) with the WR feature.
}
\label{fig:abun_ra}
\end{figure}

\begin{figure}
\begin{centering}
\includegraphics[width=1\linewidth]{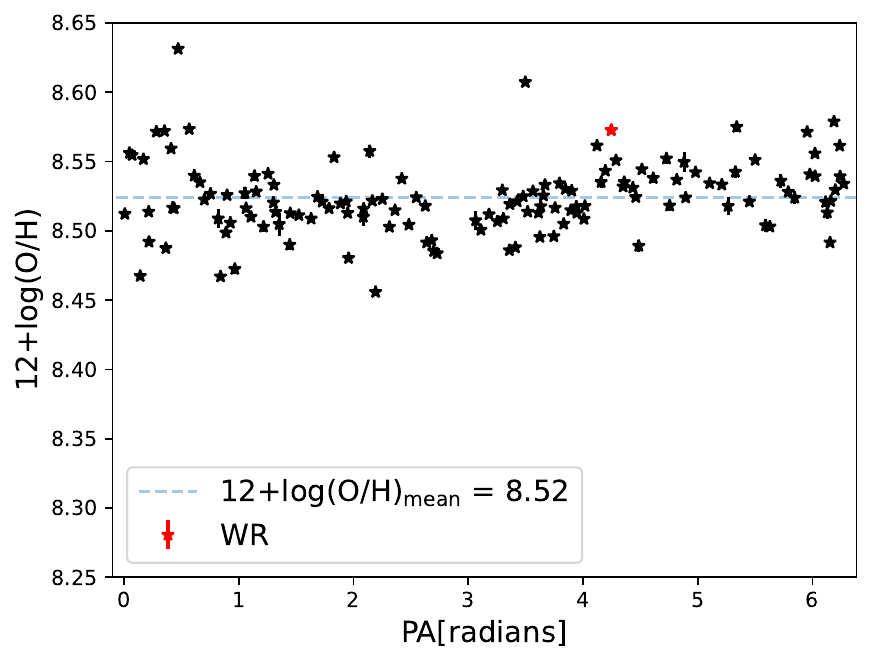}\\
\par\end{centering}
\caption{
$12+\log\rm{(\frac{O}{H})}$ vs. position angle (PA) for the \hii\ regions (black stars) in the double ring of AM\,0644-741.
O abundances were determined with the SLM.
The median value is indicated with a blue dashed line.
There is no indication of an azimuthal gradient.
The WR region is indicated by a red star.
}
\label{fig:abun_pa}
\end{figure}

We determined the abundances using the strong-line method (SLM)
for all the \hii\ regions in which we could detect \oiii, \nii\ and \sii\ lines at SNR$>3$.
As we did in the Cartwheel galaxy \citep[see][]{Zaragoza-Cardiel+2022},
we used the empirical "S-calibrator" from \citet{2016MNRAS.457.3678P},
since the strong lines used are all in the MUSE spectral range: \hb, \ha, \oiii$\lambda$4959+$\lambda$5007, \nii$\lambda$6548+$\lambda$6584 and \sii$\lambda$6717+$\lambda$6731.
Abundances could be determined for 137 regions of the sample.

In order to check for any gradient, galactocentric or azimuthal, of the O abundances in the double-ring of AM\,0644-741, we used a set of ellipses indicated in Fig.~\ref{fig:elipses} to define a galactocentric radius (R) and position angle (PA) of our catalog of \hii\ regions.
AM\,0644-741 resembles its cousin Cartwheel in the sense that the nucleus is not at the geometrical center of the star-forming elliptical ring.
We hence followed the moving-centre method adopted by \citet[][]{Marcum1992} for the Cartwheel to define the ellipses.
In this method, the center of successive ellipses is changed smoothly from the nucleus for the inner ellipses to the center of the ring of the outer ellipses.
A more detailed description of this procedure is
given in \citet[][in their Sec. 4.3]{Zaragoza-Cardiel+2022}.

In Fig.~\ref{fig:abun_ra} we plot the O abundance of the \hii\ regions vs. the galactocentric radius.
We indicate the median value of $12+\log\rm{(\frac{O}{H})}=8.52$.
By a linear fit to the L-S ring data points,
we get a slight slope indicating a small 
gradient of $-0.0012\pm0.0004$~dex/kpc. In comparison, the Cartwheel shows a  gradient of $-0.017\pm0.0001$~dex/kpc \citep{Zaragoza-Cardiel+2022} which is comparable to values found in spiral galaxies \citep{Sanchez2020}.
We find the SLM-determined O abundance of WR1 is slightly above the median value, but only by 0.05~dex.
In Fig.~\ref{fig:abun_pa} we show the O abundances vs. its position angle (PA).
Here we did not find any indication of an azimuthal gradient.
We list our results for each object in Tab.~\ref{tab:results}.
It may be noted that the O abundance of WR1 determined from the DM is  $\sim0.3$~dex higher than that determined from the SLM, which is marginally higher than the 0.2~dex error in the former method.

\section{Ionization state of the \hii\ regions}\label{Ionizing sources}

The availability of spectra covering almost the entire optical range allow us to address the ionization mechanisms of the \ha-emitting regions in the MUSE FoV.
We use \citet[][]{Xiao2018} calculations based on {\scshape BPASS} \citep[][]{Eldridge2017} and {\scshape Cloudy} \citep[][]{Ferland1998} models in standard BPT diagrams \citep[][]{Baldwin+1981} to study the ionization sources in the double ring of AM\,0644-741.
We also discuss the role of the WRs and ULXs in the ionization when \hii\ regions are associated to these sources.

\subsection{\hii\ region containing WR stars}\label{Wolf-Rayet stars}
\label{sec:wrnumbers}

WR stars are formally defined by their emission-line dominated spectra arising from their strong winds. The core-He burning objects among them, also called ``classical'' WR stars, represent a late stage in the evolution of massive stars, where a considerable amount of their outer layer has been lost.
Assuming single-star evolution, WR stars start their lives on the main sequence as O stars with $M_\text{init} \geq 25$~M$_{\odot}$
at Z$_{\odot}$ \citep[e.g.,][]{Georgy2012,Chieffi2013,Chen2015}.
See also \citet[][]{Massey2003} for estimated masses ($M_\text{init} \geq 20$~M$_{\odot}$) of the progenitor stars for different WR types, at different metallicities.
While different assumptions about mass loss or the formation of a WR stars due to binary interaction can shift the absolute mass limits, all WR stars are so massive that they usually cannot be much older than $\sim2-4$~Myr when reaching the (classical) WR stage.
Characterized by significant mass loss rates of $\dot{M}\approx10^{-4}-10^{-5}$~M$_{\odot}$~yr$^{-1}$, WR stars are expected to inject mechanical energy, and chemically processed material into the local ISM \citep[see the review by][and references therein]{Crowther2007}. With their inherently hot temperatures, classical WR stars are major sources of hydrogen ionizing photons, while their \ion{He}{ii} ionizing flux strongly depends on the wind strength and vanishes for more dense winds \citep[e.g.,][]{SanderVink2020,Sander2023,Sixtos2023}.
Given the manifold impact of WR stars on their host environment, WR stars are important engines of the cosmic matter cycle, for example in the context of ongoing star and planet formation.

\begin{figure}
\begin{center}
\includegraphics[width=0.8\linewidth]{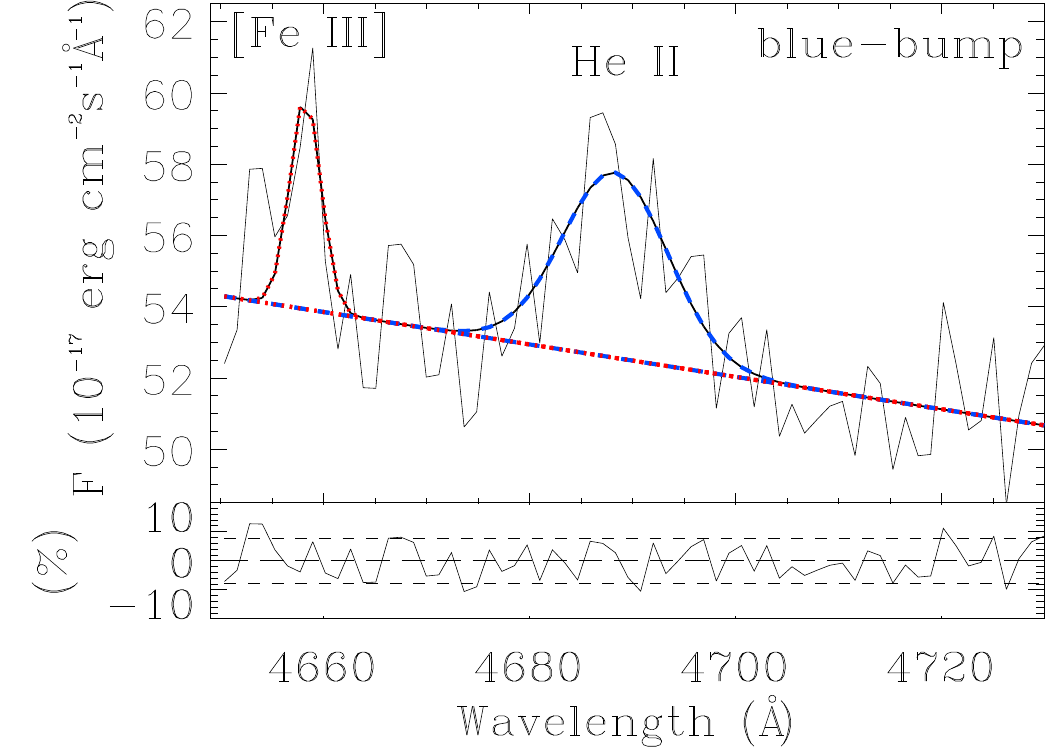}
\caption{
Gaussian fit to the BB of the \hii\ region ID\,39, WR1.
The blue dashed line represents the fit to the broad WR feature, \heiiwr, and the red dotted line the fit to the narrow nebular line.
The sum is shown in black. The fitted continuum is shown by the dashed straight line.
The residuals are shown at the bottom of the panel (in percent). The results are listed in Tab.~\ref{tab:wr}.
}
\label{fig:MUSE_spe_bb}
\end{center}
\end{figure}

For the one region where  we identified a BB (ID\,39 or WR1) in its spectrum, we can derive an approximate number of WR stars.
For this, we first measured the flux of the broad \heiiwr\ with a simple  Gaussian fitting
(see Fig.~\ref{fig:MUSE_spe_bb}).
Similar to the study of \citet[][in NGC\,1569]{Mayya2020}, the line of \feiiia\ is also present in emission, but it is well separated from the BB, so there is no need for a de-blending procedure, which was necessary in \citet[][]{Gomez-Gonzalez+2021}.
Given a distance of 98.62~Mpc, we obtain the luminosity of the \heii\ line, assuming it dominates the BB: $L$(\heii-broad)$=7.78\times10^{38}$ \ergs.
We do not detect \civr, a.k.a. the `red-bump' (RB), ruling out a large presence of WC or WO-type stars.
This also justifies our %calculation 
assignment of the entire fitted flux as that of the broad \heii\ line
of $L$(\heii) as WC stars could contribute to the BB with a \ciiib\ line, sometimes even exceeding the intensity of \heii. However, such BBs tend to display a broader FWHM \citep[see, e.g.,][]{Gomez-Gonzalez+2020} which is not the case for WR1.
Thus we conclude that WR1 contains mainly WN stars and calculate the number of
a typical Nitrogen-rich late-type WNL star with the
expression given by \citet[][]{Lopez-SanchezEsteban2010}:
L$_\text{WNL}$(\heii-broad) = $(-5.430+0.812\text{x})\times10^{36}$~\ergs
where $\text{x}$ is the O abundance.
Using the value obtained by the DM, $\text{x}=8.9$, we get
$L_\text{WNL}$(\heii) = $1.472\times10^{36}$~\ergs.
From the observed $L$(\heii)$=7.78\times10^{38}$ \ergs, we estimate a number of 430 WR stars in this region.
The aperture diameter of the extracted spectrum is $D_{\rm aper}$=2.324~arcsec, which corresponds to a size of 1.1\,kpc.
Our results are shown in Tab.~\ref{tab:wr_results}.

\begin{table}
\small\addtolength{\tabcolsep}{-0.5pt}
\begin{center}
\caption[]{Observed and derived parameters of WR1.
}
\label{tab:wr_results}
\begin{tabular}{cc}
\hline
Parameter                          & value  \\
\hline
L(\heii)                           & $7.78\pm0.39\times10^{38}$ \ergs\  \\
FWHM(\heii)                        & 12.1~\AA            \\ 
EW(\heii)                          & 1.3~\AA            \\
L(\hb)                             & $3.15\pm0.05\times10^{40}$ \ergs\  \\
N$_\text{WNL}$ ($12+\log\rm{(\frac{O}{H})}=8.9$)   & $430\pm20$     \\
N$_\text{O7V}$                     & 6620$\pm$100            \\ 
WR/O                               & 0.065$\pm$0.003   \\ 
\hline

\end{tabular}
\\
Assumed values from \citet[][]{Lopez-SanchezEsteban2010}:
 L$_\text{WNL}$(\heii) = $1.80\times10^{36}$~\ergs ($12+\log\rm{(\frac{O}{H})}=8.9$);\\
 L$_\text{O7V}$(\hb) = $4.76\times10^{36}$~\ergs. \\
\vspace{0.4cm}
\label{tab:wr}
\end{center}
\end{table}

By using fluxes from optical spectrophotometry and distances determined with recently available Gaia DR3 parallaxes,
\citet[][]{Crowther2023} report line luminosities for MW, Large Magellanic Clouds (LMC) and Small Magellanic Clouds (SMC) WR stars.
Prior to this work, the expression from \citet[][]{Lopez-SanchezEsteban2010} was commonly used to quantify the WNL stars required for the observed \heii\ luminosity \citep[e.g.][]{Miralles2016,Mayya2020} in WR galaxies \citep[defined by][]{Conti1991}.
\citet[][]{Crowther2023} provides detailed information for the subtypes of each classification. In their Tab. 2 they report \heii\ luminosities for WN2-5w, WN3-7s, WN6-8, WN9-11, WN5-7h and Of/WN that go from $0.5-2\times10^{36}$~\ergs, with uncertainties of the same order, at MW metallicity.
Given the distance of our object and the quality of our spectrum (SNR, spatial resolution), we can not provide a detailed sub-classification for the WN stars in WR1, since \heii\ is the only broad line detected.
However, if we consider the values by \citet[][]{Crowther2023}
the number of WN stars in WR1 is the exact value depending on the WN subtype which is in the range between 400 and 1500 WR stars, consistent with the value obtained using the expression from \citet[][]{Lopez-SanchezEsteban2010}.

We further estimate the number of O-type stars in the \hii\ regions by
assuming a ``typical'' luminosity of L(\hb)=$4.76\times10^{36}$~\ergs
emitted by O\,7V stars \citep[][]{Lopez-SanchezEsteban2010}.
Considering a derived number of 6620 O-type stars in WR1, we obtain a WR/O-ratio of $0.1$, which would be consistent with the time that O-type stars are expected to later spend in their WR phase, namely typically $\sim10\%$ of their lifetime \citep[see][]{Crowther2007}.

\subsection{ULX sources}\label{ULX sources}

ULXs are point X-ray sources located off-center in their host galaxies and having 
X-ray luminosities
$L_{\rm X} > 3\times 10^{39}$\,erg\,s$^{-1}$, i.e.\ exceeding the
Eddington luminosity of an accreting black hole with $M_\bullet \approx 20\,M_\odot$ 
\citep{Kaaret2017}. Most likely, ULXs are binaries consisting of a massive donor star and 
a neutron star or a few $M_\odot$ black hole, accreting supercritically \citep{Poutanen2007,Stobbart2006,Gladstone2009,Sutton2013}.
However, stellar evolutionary models predict that some ULXs indeed may contain black holes with 
masses $> 10\,M_\odot$ \citep[e.g.][]{Finke2017,
Marchant2017, Hainich2018}. These ULXs would
represent an advanced evolutionary stage of a very massive binary star, and therefore are 
expected to be associated with young massive clusters. Indeed, it is not uncommon to find ULXs 
located in the vicinity of massive star clusters \citep[][]{Berghea2013,Poutanen2013,Oskinova2019}.

\begin{table}
\small\addtolength{\tabcolsep}{-0.5pt}
\begin{center}
\caption{X-ray sources and their luminosities reported in AM\,0644-741 by \citet[][assuming a distance of 91.6 Mpc from NED in 2018; updated values assuming 98.62~Mpc (NED) are shown in parentheses]{Wolter+2018}${^\dagger}$ and their nearest \hii\ region analyzed in this work${^\dagger}{^\dagger}$.
Except for W3 (nucleus) and W5 (AGN${^a}$), these objects have enough luminosities ($L_{X}>1\times10^{39}$ \ergs) to be considered ULX sources.}
\setlength{\tabcolsep}{0.3\tabcolsep}
\begin{tabular}{cccccc}
\hline
ID${^\dagger}$ & $L_{X}$[0.5-10~keV]${^\dagger}$ & $L_{X}$[2-10~keV]${^\dagger}$ & Nearest${^\dagger}{^\dagger}$      & \multicolumn{2}{c}{offset} \\
   & ($\times10^{39}$ \ergs) & ($\times10^{39}$ \ergs)   & \hii\ region & (arcsec)  & (pc) \\
\hline
W1 &  8.0 (9.3) & 5.1 (5.9)  & WR1$^{*}$ & 0.75 & 358.6 \\
W2       & 12.6 (14.6)& 8.1 (9.4)  & ID\,13            & 2.0        & 956.2 \\
W3       & 28.6 (33.2)&18.1 (21.0) & AGN${^a}$     & \dots        & \dots \\
W4 &  9.8 (11.4)& 6.3 (7.3)  & ID\,92$^{*}$      & 1.1  & 525.9 \\
W5       & 22.4 (26.0)&14.3 (16.6) & nucleus        & \dots        & \dots \\
W6       &  2.1 (2.4) & 1.3 (1.5)  & ID\,158           & 2.1        & 1004.1\\
W7       &  2.3 (2.7) & 1.4 (1.6)  & ID\,170           & 1.7        & 812.8 \\
W8       &  5.1 (5.9) & 3.3 (3.8)  & ID\,113           & 2.85       & 1362.7 \\
W9 &  8.5 (9.9) & 5.4 (6.3)  & ID\,142$^{*}$     & 0.8  & 382.5 \\
\hline
\end{tabular}\\
(${a}$) \citet[][]{Heida+2013} and corroborated in this work; 
(${*}$) probably related with the nearby \hii\ region; projected location $<1$~arcsec.
See text for details.
\label{tab:ulx}
\end{center}
\end{table}

The AM\,0644-741 galaxy presents an interesting case for studies of ULXs populations.  
Nine X-ray sources have been detected in this galaxy \citep[][see their Fig.~\ref{fig:muse_image}]{Wolter+2018}. Among them, one is the nucleus of  AM\,0644-741, and 
another one is a 
background AGN \citep[][]{Heida+2013} (see also Appendix B). Seven remaining X-ray sources are ULXs. Three of them
(W1, W4, and W9) are located at $<1.1$~arcsec in projection from star clusters and their \hii\ 
regions (WR1, 92, and 142). The other three ULXs are also located within the star-forming ring and are likely
associated with nearby \hii\ regions. Only one ULX, W8, located in the NE section of the ring, has no obvious optical counterpart \hii\ region or star cluster. The projected separation of the X-ray sources to their nearest \hii\ regions are listed in Tab.~\ref{tab:ulx}.

The ULX W1 is especially interesting since its position coincides with the brightest \hii\ region 
in its host galaxy surrounding a cluster of WR stars (Figs.~\ref{fig:MUSE_spec}). 
Given the age and the mass of the associated star cluster, we tentatively suggest that W1 is powered by 
the accretion onto a black hole.
Several ULXs have been observed in association with He\,{\sc iii} regions \citep{Pakull2003}.
However, not all ULXs are associated with a nebular region \citep[][]{Feng2011}.
For example in the Cartwheel only 10 of the 17 ULX sources have associated \heii\ emission.
However, line ratios of commonly observed nebular lines do not show any signature of the ionization by X-ray sources \citep[][]{Mayya2023}.
No nebular He\,{\sc ii} lines are detected in
the MUSE spectra of the ULX in AM\,0644-741.
This implies that the presence of a ULX and large numbers of WR stars are not a sufficient condition for producing a He\,{\sc iii} nebula as discussed recently by e.g.\ \citet{Schaerer2019} and \citet{Umeda2022}.
Given the higher than solar metallicity of AM\,0644-741,
the winds of the WR stars are likely too dense to provide a
sufficient budget of He\,{\sc ii} ionizing photons.
Recently, \citet{Oskinova2022} suggested that the lower temperature of superbubbles around lower metallicity star clusters could help to explain the observed prevalence of
He\,{\sc iii} nebulae in low-metallicity galaxies. 

Another interesting source is W3, the brightest X-ray source on the FoV of AM\,0644-741.
\citet[][]{Heida+2013} identified this object as an AGN due to detecting \mgiia\ and \ciiip\ with the VLT/FOcal Reducer and low dispersion Spectrograph (FORS2) observations. In Fig.~\ref{fig:qso} we show the MUSE spectrum extracted at the position of W3, identified as an \ha-emission region in our first SExtractor catalog. This mis-identification happened due to the \mgiia\ line of the AGN being redshifted precisely to the wavelength of 
\ha, which was the criteria to construct our catalog. Additionally, we identify \oiia\ and \neiiia\ at a similar redshift of $z \sim 1.4$. Their redshifted wavelength range was not covered in previous observations. Thus, we independently confirm that W3 is a 
background AGN at $z \sim 1.4$.
On top of the wide emission line, we identify a narrow \ha\ absorption line at the redshift of AM\,0644-741 (z$\sim$0.022).
We use BPT diagrams to check whether any of the observed lines of regions
containing ULX sources require ionization from X-rays.

\subsection{BPT diagrams}\label{BPT diagrams}

\begin{figure*}
\begin{centering}
\includegraphics[width=0.48\linewidth]{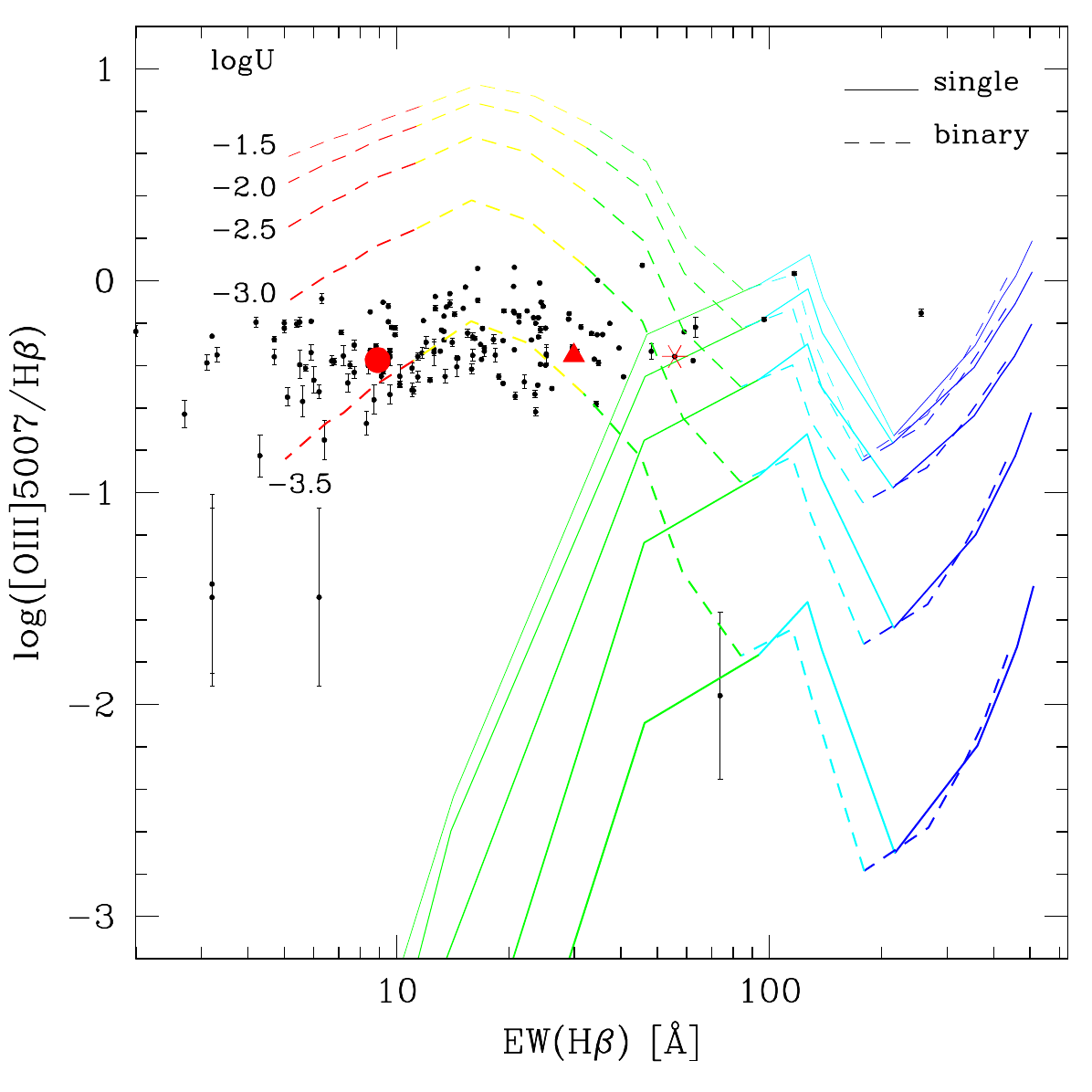}
\includegraphics[width=0.48\linewidth]{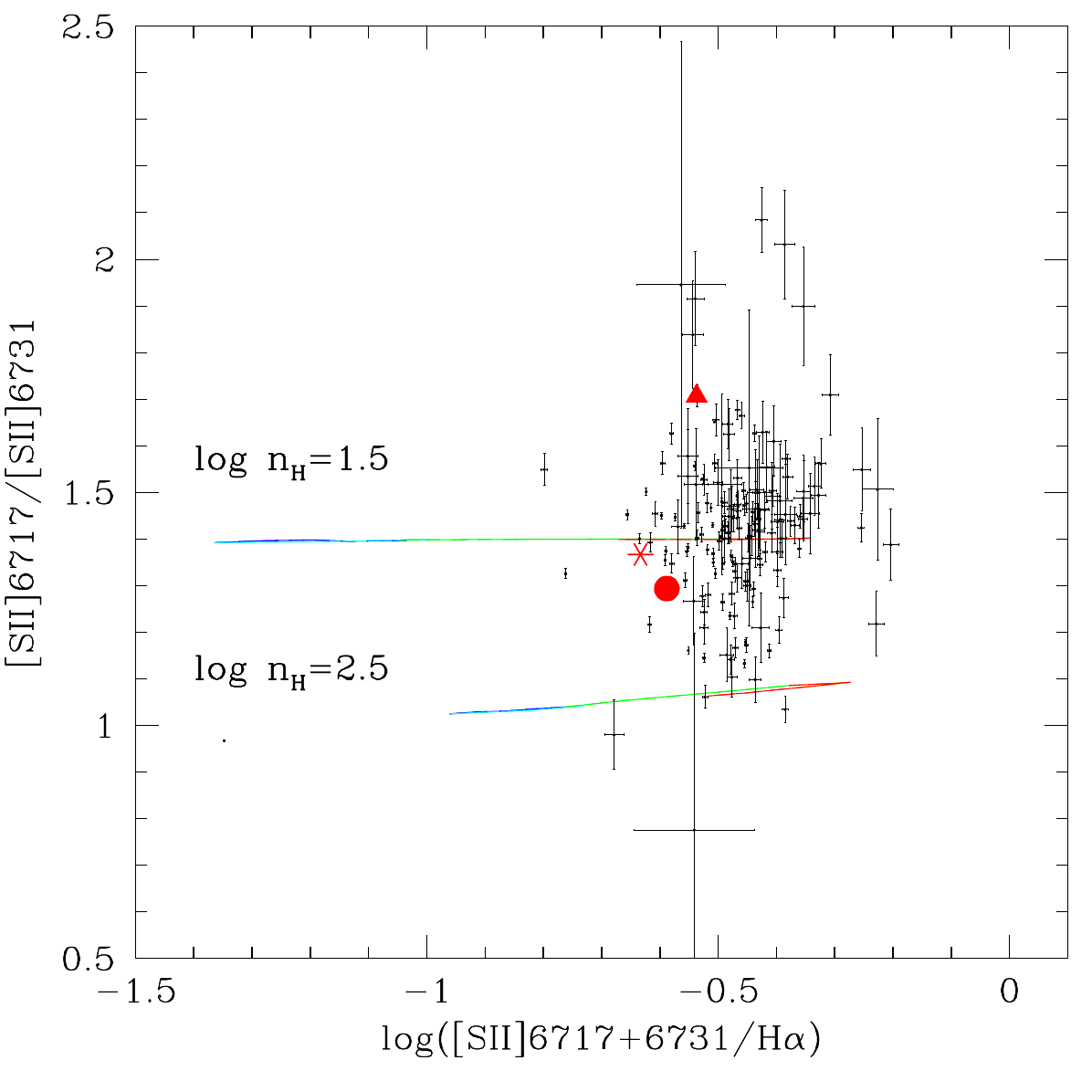}
\par\end{centering}
\caption{
Diagrams for the diagnostic of age and ionization state (left) and density (right) of the \hii\ regions in AM\,0644-741 (black points with vertical error bars). Red points correspond to special \hii\ regions that have ULX emission (star ID\,39=WR1, triangle ID\,92 and circle ID\,142). (left) Photoionization model results from \citet{Xiao2018} for {\scshape BPASS} single (continuous line) and binary (dashed line) models are plotted at Z=0.017, density $\log n_{\rm H}[{\rm cm}^{-3}]$=2 for five values of $\log U$, which are identified by the labels. The curves are colour-coded to denote the ages: 1--2.5~Myr blue, 2.5--5~Myr cyan, 5--10~Myr green, 10--20~Myr yellow and 20--100~Myr red. 
(right) Colour coding of the models is the same as on the left except that only the binary models at $\log$U=$-$3.5 are shown for two densities which are indicated. Binary low ionization models ($-3.5<\log U<-3.0$) at ages $>$5~Myr at $n_\mathrm{e}$ of 100~cm$^{-3}$ fit the four plotted quantities simultaneously for the majority of the regions.
}
\label{fig:oiii-ew}
\end{figure*}

\begin{figure*}
\begin{centering}
\includegraphics[width=0.48\linewidth]{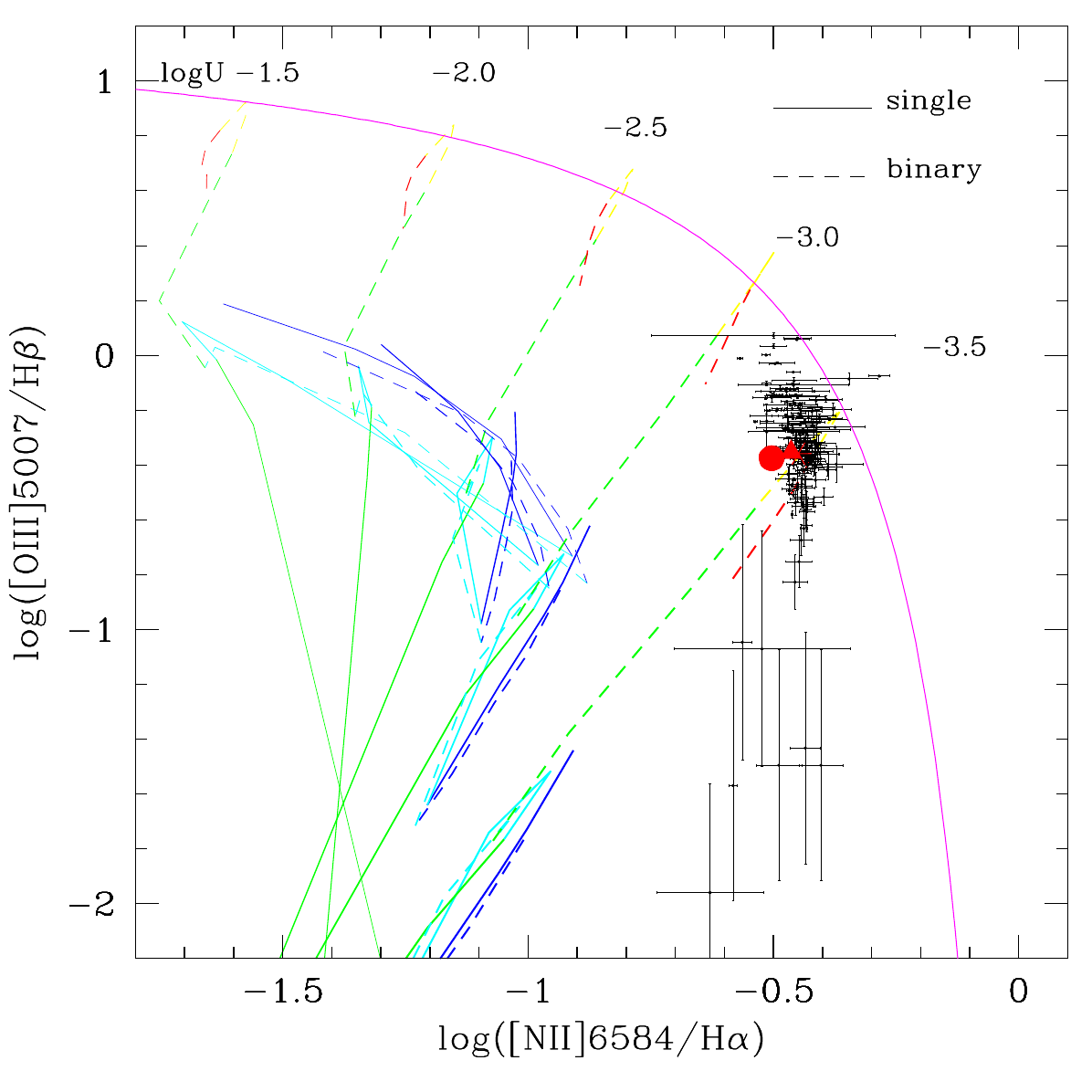}
\includegraphics[width=0.48\linewidth]{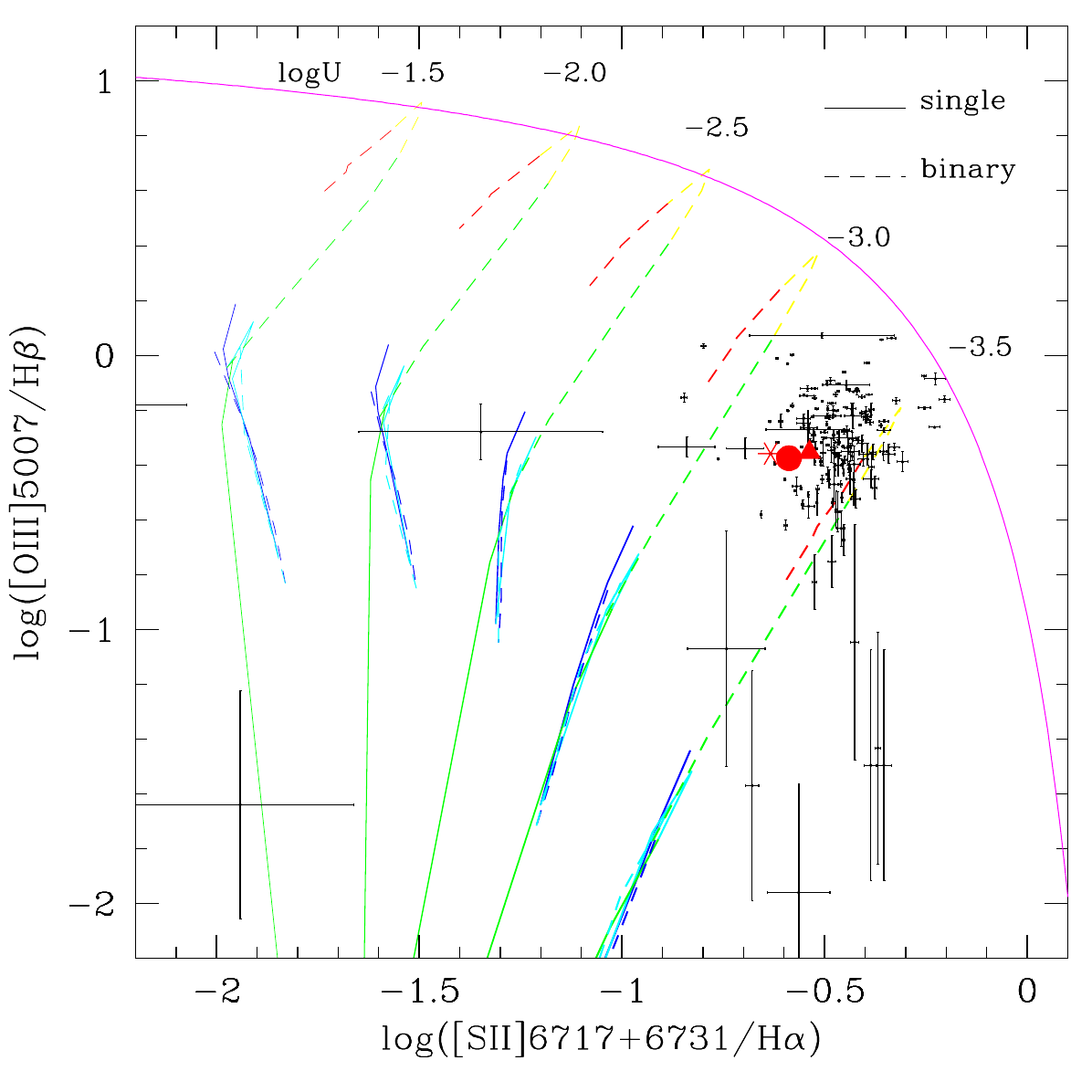}
\includegraphics[width=0.48\linewidth]{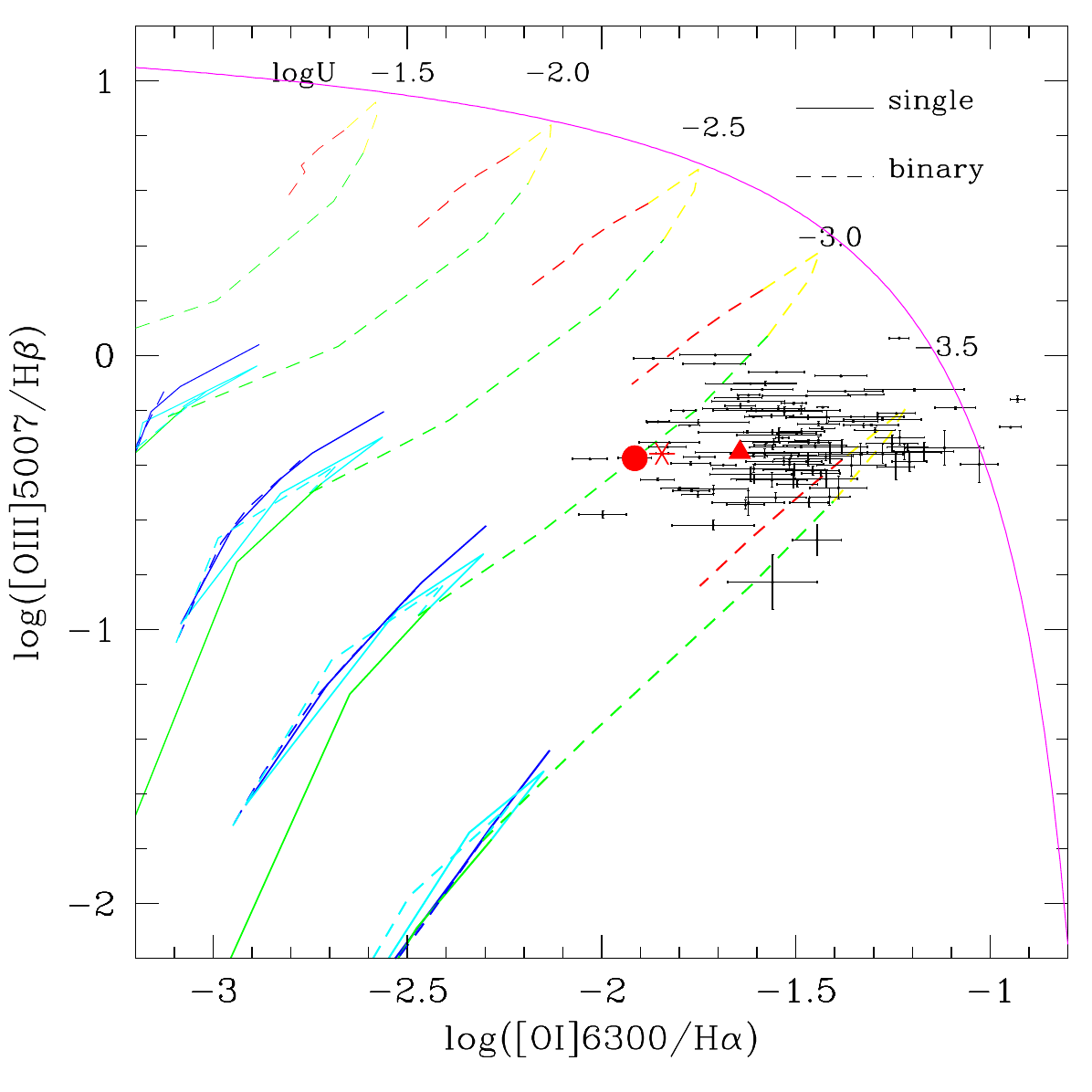}\\
\par\end{centering}
\caption{
AM\,0644-741 \hii\ regions in diagnostic BPT \citep[][]{Baldwin+1981} diagrams; (upper left) \oiiibyhb\ vs. \niibyha\ line ratio;
(upper right) \oiiibyhb\ vs. \siibyha;
(bottom) \oiiibyhb\ ratio vs. \oibyha. 
Notations and colour-coding of the models and observed points are the same as in Fig.~\ref{fig:oiii-ew}.
Regions photoionized by stars are expected to lie below the magenta curve defined by \citet{Kewley2006}.
See text for details.
}
\label{fig:bpt1}
\end{figure*}

We use the standard diagnostic diagrams to determine the physical parameters such as the age, ionization parameter (\logU) and gas density ($\log n_{\rm H}$) for the \hii\ regions in the double ring of AM\,0644-741. For this purpose we use the photoionization calculations of \citet[][]{Xiao2018} which are based on {\scshape Cloudy} models
\citep[][]{Ferland2017}
using single and binary population synthesis models of {\scshape BPASS} \citep{Eldridge2017}. 
We found that the observed points lie between the calculated values for metallicities Z$=0.014$ and 0.020. We hence generated a new set of tables corresponding to Z$=0.017$ by interpolating the fluxes of each nebular line from the Z$=0.014$ and 0.020 tables. These tables were generated for both single and binary star models.

The EW(\hb) is a parameter commonly used as an indicator of the age, whereas  \oiiibyhb, \niibyha, and \siibyha\ are sensitive to the ionization state of a starburst region. Flux ratios of the \sii\ doublet are sensitive to the $n_\mathrm{e}$ of \hii\ regions. We hence compare the observed values of these quantities with the results from photoionization models. 
In Fig.~\ref{fig:oiii-ew}, we plot the \oiiibyhb\ line ratio vs. \ewhb\ on the left and \sii\ doublet line ratios vs \siibyha\ on the right. Errors on observed quantities are plotted in both axis but is noticeable only when it is larger than the symbol size. The three \hii\ regions containing ULX sources, one of which is also a WR source, are shown by large red coloured symbols.

On the left figure, photoionization models are shown for single and binary models with continuous and dashed lines, respectively, at five values of $\log$U ranging from $-3.5$ to $-1.5$ at an interval of 0.5~dex. Along each curve,  age of the ionizing cluster changes from 1~Myr to 100~Myr, with the colour of the curve denoting interesting ranges of ages: 1--2.5~Myr (blue), 2.5--5~Myr (cyan), 5--10~Myr (green), 10--20~Myr (yellow) and 20--100~Myr (red). The plotted models correspond to density $\log n_{\rm H}[{\rm cm}^{-3}]$=2.

On the right Figure, we show the binary evolutionary tracks for $\log n_{\rm H}[{\rm cm}^{-3}]$=1.5 and 2.5. The plotted \sii\ doublet ratio becomes insensitive to density for $n_{\rm H}\lesssim 30~{\rm cm}^{-3}$, whereas the ratio decreases for densities higher than the plotted limit. The Figure illustrates that AM\,0644-741 regions lie between the two plotted ranges, and hence we take a mean density of $n_{\rm H}=100~{\rm cm}^{-3}$.

Binary models at low ionization  ($-3.5<\log U<-3.0$) at ages $>$5~Myr at $n_\mathrm{e}$ of 100~cm$^{-3}$  fit the four plotted quantities simultaneously for the majority of the regions.
We derive an age of $\sim$6~Myr with or without binary for WR1, which is also a ULX (red asterix), whereas for the other two ULX sources, we derive ages of 10 and 25~Myr respectively for ID\,92 and ID\,142. Only binary models reproduce all the observed quantities for these two ULX sources.

In Fig.~\ref{fig:bpt1} we plot the observed \hii\ regions in the three standard BPT diagrams \citep[][]{Baldwin+1981}: 1) \oiiibyhb\ vs. \niibyha, 2) \oiiibyhb\ vs. \siibyha\ and 3) \oiiibyhb\ vs. \oibyha\ line ratios. 
The same photoionization models that are used in the left panel of  Fig.~\ref{fig:oiii-ew} are shown on these diagrams.
In addition, we draw the maximal starburst locus of \citet{Kewley2006} in these plots.
As expected, all the \hii\ regions are located in the star-forming zone, including those containing ULX and WRs.
In fact, the location of these ULX sources does not differ from that of the rest of the \hii\ regions in these BPT diagrams. 
The X-ray photons from ULX sources are expected to ionize the neutral medium surrounding the \hii\ regions, creating energetic free electrons in the cold medium. These free electrons collisionally excite oxygen in the cold neutral medium, thus increasing the \oibyha\ ratio to values above those for normal \hii\ regions. \citep{Gurpide2022}. No such effect is seen, and hence the ULX sources do not contribute in any significant way to the ionization of \hii\ regions.
It can be noticed that the same low ionization  ($-3.5<\log U<-3.0$) binary models at ages $>$5~Myr and $n_\mathrm{e}$ of 100~cm$^{-3}$ that best fit the \oiiibyhb\ vs \ewhb\ diagram also fit the plotted quantities simultaneously for the majority of the regions.

As mentioned in the beginning of this section, we have used the results of the photoionization code for Z=0.017. This value 
corresponds to a gas phase metallicity of $12+\log(\frac{\rm O}{\rm H})=8.88$ for dust-free \hii\ regions, whereas it would be 8.65 for dust-to-metal mass ratio of $\xi_{\rm d}$=0.5 \citep{Gutkin2016}.
The inferred gas phase metallicity is closer to the metallicity obtained from the DM rather than the SLM. 
At lower metallicities the \oiiibyhb\ line ratio requires a higher $\log$U as compared to the \niibyha\ line ratio. 

Single star evolutionary models in {\scshape BPASS} SSP code are able to reproduce the observed values only for regions whose \ewhb$>$50~\AA, which corresponds to an age of $<$6~Myr. At older ages, line ratios fall rapidly for single star models, whereas the ratios are marginally higher for binary star models. This trend seen in AM\,0644-741 is similar to the results obtained by \citet[][]{Xiao2018} for the sample of \hii\ regions used in their study.

\section{Discussion}\label{Discussion}

\subsection{Abundances}\label{Abundances}

Until now, there has not been a comprehensive study of the chemical abundances in AM\,0644-741.
The first reports in this regard were made by
\citet[][]{Higdon2011}, who reported an O abundance of $12+\log\rm{(\frac{O}{H})}=9.1$ for three regions in the ring of AM\,0644-741.
The locations of their slitlets correspond to our objects identified as ID\,84 (their slitlet 1); ID\,44, 42, 39 and 37 (slitlet 2); and ID\,9 (slitlet 3).
They used EFOSC2 spectra covering the spectral range from 6200 to 7800~\AA.
Given the wavelength range limitations of their spectra, they obtained these values based on F(\nii)/F(\ha) and F(\nii)/F(\sii) relations from \citet[][]{Nagao2006}.
Actually, \citet[][]{Higdon2011} observed the 3500 to 5500~\AA\ wavelength range, but due to problems in the reduction process did not use these spectra in their analysis, which is probably the reason why they did not report the
WR feature at 4686~\AA\ in their slitlet 2, where the brightest \hii\ region ID\,39 (WR1) is located.

Given the FoV and wavelength range covered by MUSE observations we are able to determine O abundances for 137 regions in the ring of AM\,0644-741, including those covered by \citet[][]{Higdon2011}. 
We report a median value of $12+\log\rm{(\frac{O}{H})=8.52}$, i.e. slightly subsolar. However, caution should be exercised when comparing this value obtained with the S-calibrator from \citet{2016MNRAS.457.3678P} to those values obtained with different SLM calibrators, as differences of 0.5~dex are common. Conversions between abundances from different SLM calibrators up to the precision of less than 0.1~dex for most of the SLM calibrators do exist \citep{2022MNRAS.512.3436E}. We also obtain a slight gradient of
$-0.0012\pm0.0004$~dex/kpc among the ellipses 7 and 11. 
With this estimation we were able to compare with proper ionizing models in BPT diagrams and
to determine the number of WR stars in WR1. Using the DM, we obtain a supersolar value with $12+\log\rm{(\frac{O}{H})=8.9\pm0.2}$, albeit already 0.2~dex lower than previously reported by \citet[][]{Higdon2011}.

\subsection{Comparison with the Cartwheel}\label{Comparison with the Cartwheel}

AM\,0035-335 (the Cartwheel) and AM\,0644-741, are probably the most iconic and studied ring-galaxies in the literature.
This is in part explained by their striking ring morphology and also their relative proximity, located at 128~Mpc and 98.6~Mpc, respectively.
In \citet[][]{Zaragoza-Cardiel+2022} and \citet[][]{Mayya2023} we started a series of studies in ring-type galaxies with available MUSE data to analyse the star-formation triggered in the ring of these peculiar of galaxies.
In \citet[][]{Zaragoza-Cardiel+2022} we reported that the \hii\ regions in the ring of the Cartwheel have a median O abundance of $12+\log\rm{\frac{O}{H}}$=8.19$\pm$0.15, slightly higher than in the SMC.
Using the SLM, we here report a median value of $12+\log\rm{\frac{O}{H}}$=8.52 for AM\,0644-741, clearly above Cartwheel and in fact slightly higher than the LMC average \citep[e.g.][]{ToribioSanCipriano2017}.

Comparing the BPT diagram locations of the \hii\ regions studied in this work with those in Cartwheel by \citet[][]{Mayya2023}, it is evident that the objects in AM\,0644-741 are in a considerably lower state of ionization.
Looking at the \oiiibyhb\ vs. \niibyha\ diagram, for example, the \hii\ regions in Cartwheel are located in the range $0.2<$\oiiibyhb$<0.9$ and $-1.6$\niibyha$<-1.0$ while regions in AM\,0644-741 are between $-0.8<$\oiiibyhb$<0.0$ and $-0.5$\niibyha$<-0.6$.
The same can be seen in the other diagrams for \oiiibyhb\ vs. \siibyha\ and \oiiibyhb\ vs. \oibyha,
our objects are in another zone, at the bottom left of the BPT diagrams relative to the Cartwheel regions, where the ionization parameter can be as high as \logU$=-2.0$, an order of magnitude higher than in AM\,0644-741.

Also, \citet[][]{Mayya2023} report the detection of nebular \heiiwr\ in 32 \hii\ regions in the Cartwheel, with a mean I(\heiiwr)/I(\hb) ratio of 0.010$\pm$0.003. In contrast, we do not detect nebular \heiiwr\ in the L-S ring.
Instead, we detect the broad \heiiwr\ in the spectrum of a knot of star-formation with a luminosity 
consistent with the presence of at least $\sim$430 WR stars.
The non-detection of nebular \heiiwr\ in spite of the detection of a WR feature in the L-S ring is likely due to metallicity difference. At the slightly higher metallicity of L-S ring, it is expected that the atmosphere absorbs the He+ ionizing photons more efficiently than at lower metallicity of the Cartwheel.

Like the Cartwheel, AM\,0644-741 contains a relatively high number of ULXs, conforming the 
universal scaling of the X-ray luminosity function and star formation rate in galaxies \citep[SFR,][]{Mineo2012}.
According to \citet[][and references therein]{Wolter+2018},
Cartwheel with a SFR of 20~$M_{\odot}$/yr has 15~ULXs and
AM\,0644-741 with a SFR of 11.2~$M_{\odot}$/yr has 7~ULX sources.
An anti-correlation
between the number of ULXs per host galaxy normalized to the SFR
vs. the metallicity of the host galaxy is empirically established \citep[][and references therein]{Wolter+2018, Lehmer2021}.
The number of ULXs normalized to the SFR in this galaxy is consistent with the metallicity we report in this work.
However, we do not find any evidence of additional ionization by X-ray sources in the \hii\ regions containing ULX sources in AM\,0644-741, a result which is similar to that found by \citet[][]{Mayya2023} for the Cartwheel \hii\ regions.

\section{Summary and concluding remarks}\label{Summary and concluding remarks}

We have presented an analysis of a sample of \hii\ regions
in the star-forming double-ring of the AM\,0644-741 collisional galaxy using spectra we extracted from available VLT MUSE observations.
We constructed a new catalogue of 179 objects following a semi-automatic approach with the standard tool SExtractor based on the \ha\ emission of these sources.
Until now, this is the largest sample of \hii\ regions analyzed in this galaxy. This sample could serve as a reference for future studies on star formation triggered in the rings of post-collisional galaxies.
After a proper extraction, dereddening, correction for underlying absorption using SSP synthesis models and the flux measurement of all the emission lines present in the spectra, we investigated the chemical abundances and ionization state of these sources.
Here we summarize our findings:

\begin{itemize}

\item  We determined the O abundances for 137 \hii\ regions using the SLM.
The rest of the objects did not have all the necessary lines to determine their abundance.
We report a median a median O abundance value of $12+\log\rm{(\frac{O}{H})}=8.52$.
This value is between solar \citep[$12+\log\rm{(\frac{O}{H})}_\odot=8.69$;][]{Asplund2009}
and LMC \citep[$12+\log\rm{(\frac{O}{H})}\sim8.4$;][and references therein]{Stasinska2012}.
A linear fit  to the data points among the ellipses including the \hii\ regions of the ring indicates a slight gradient of $-0.0012\pm0.0004$~dex/kpc. There is no indication of an azimuthal gradient.
This information was used to select appropriate photoionization models based on the metallicity for further analysis of the physical conditions of \hii\ regions in this galaxy.

\item We calculated ionic abundances of He, N, O and Fe using the DM for the brightest \hii\ region, ID\,39. Only this region has all the necessary lines to determine chemical composition via this method.
We find values of
$12+\log\rm{(\frac{He}{H})}=10.75\pm0.14$;
$\log\rm{(\frac{N}{O})}=-1.3\pm0.2$;
$12+\log\rm{(\frac{O}{H})}=8.9\pm0.2$ and
$\log\rm{(\frac{Fe}{O})}=-2.7\pm0.3$.
We also determine other physical parameters of the \hii\ regions in the double-ring of this galaxy like flux, SNR and EW of H$\beta$, visual extinction \Av\ and report updated radial velocities.

\item  We found a WR feature, the BB, in the brightest \hii\ region, naming it WR1.
The luminosity of the observed \heiiwr\ broad line is consistent with the presence of at least $\sim$400 WNL WR stars 
in this knot of massive star-formation.
We have carried out a careful search for nebular \heiiwr\ in all the \hii\ regions of this galaxy, but do not detect any narrow \heiiwr\ in any of the analyzed regions.

\item We estimate the number of O-type stars in the \hii\ regions
assuming a typical luminosity emitted by O\,7V stars.
The $L$(\hb) in WR1 in particular, is consistent with a number of 6620 O-type stars, giving a WR/O ratio of $0.1$.
This is consistent with the length of the WR phase, $\sim10\%$ of the lifetime of the massive star progenitor.

\item We observe that the location of at least three of our objects (WR1, 92 and 142) coincide with of one of the ULX sources reported by \citet[][]{Wolter+2018} in this galaxy (W1, W4 and W9).
The ULXs are related to the star-forming regions in the ring \citep[][]{Wolter+2018}.
However, we encountered that the exact \hii\ region in our sample is not very clear for half of the ULX sources.
Other three ULX sources (W2, W3 and W8) are located a projected distance $>1$~arcsec from the nearest \hii\ regions (ID\,13, 158 and 113).
The analysis of similar ring systems with the presence of ULXs could help to settle their nature and the role they play in the ionization of star-forming regions.
However, the optical spectrum does not show line ratios characteristic of ionization by an X-ray source.

\item We independently discard the ULX nature of the W3 \citep[][]{Wolter+2018},
reporting two additional lines from those reported by \citet[][]{Heida+2013} at $z \sim 1.4$ by using novel spectra covering a longer wavelength range.
Our result supports the AGN nature of W3.
Moreover, we show that its location coincides with a star-like counterpart in optical bands using available HST images.

\item By using standard optical diagnostic diagrams with the available emission lines in our spectra: \oiiibyhb\ vs. \siibyha\ and \oiiibyhb\ vs. \oibyha, we find that our \hii\ regions are consistent with photoionization by star clusters with ages of 2.5-20~Myr. Comparing the observed line-ratios with single and binary evolution SSP models, we estimate $-3.5<$\logU$<-2.9$ and $\log\langle n_{\rm H} \rangle$.
In three BPT diagrams, a binary population is needed to reproduce the observations.

\item In this work we have compared the star-forming regions in the two collisional ring galaxies observed so far with VLT MUSE: the Cartwheel and AM\,0644-741.
They are similar in morphology but turn out to be different in metallicity.
The Cartwheel is a relatively metal-poor galaxy \citep[][]{Zaragoza-Cardiel+2022} with a single external ring while AM\,0644-741 is a double-ring metal-rich galaxy.

\item In the Cartwheel, \heii\ nebular has been reported by \citet[][]{Mayya2023}. In contrast, AM\,0644-741 does not show any nebular \heii\, but broad \heii\ in one region. These findings are consistent with the metallicities determined for the two galaxies.

\end{itemize}

\section*{Acknowledgements}

We thank the anonymous referee for their critical reading and valuable suggestions that improved the presentation of the paper.
This work is based on data obtained from
the ESO Science Archive Facility, program ID: 106.2155.
Observations made with the NASA/ESA {\it Hubble Space Telescope} were obtained
from the data archive at the Space Telescope Science Institute.
STScI is operated by the Association of Universities for Research in
Astronomy, Inc. under NASA contract NAS 5-26555.
This research has made use of the NASA/IPAC Extragalactic Database (NED), which is funded by the National Aeronautics and Space Administration and operated by the California Institute of Technology.
VMAGG is funded by the Deutsche Forschungsgemeinschaft (DFG - German Research Foundation), grant number 443790621.
YDM thanks CONACYT (Mexico) for the research grant CB-A1-S-25070.
AACS is funded by the Deutsche Forschungsgemeinschaft (DFG - German Research Foundation) in the form of an Emmy Noether Research Group -- Project-ID 445674056 (SA4064/1-1, PI Sander). AACS further acknowledges supported by funding from the Federal Ministry of Education and Research (BMBF) and the Baden-W{\"u}rttemberg Ministry of Science as part of the Excellence Strategy of the German Federal and State Governments.

\section*{Data availability}\label{Data availability}

Fluxes of the emission lines used in this work are available
in its online supplementary material. The reduced fits files
on which these data are based will be shared on reasonable request.
The ESO datacubes are in the public domain. The link and observation ID are available in the article.

%%%%%%%%%%%%%%%%%%%% REFERENCES %%%%%%%%%%%%%%%%%%

\bibliographystyle{mnras}
\bibliography{references} 

\appendix

\section{Physical parameters}
\label{Physical parameters of the sample}
In Tab.~\ref{tab:results} we report our catalogue the 179 \hii\ regions in the  double-ringed AM\,0644-741 galaxy.
Coordinates and physical parameters like the flux log(F(H$\beta$)), SNR, EW, visual extinction \Av, number of O-type stars N$_\text{O7V}$, radial velocity $V_{rad}$ and O abundances $12+\log\rm{(\frac{O}{H})}$ are listed. Nine out of the 179 sources with \ha\ emission does not have \hb\ with SNR$>3$ and therefore only their coordinates are reported. Deeper observations could improve the SNR of the fainter \hii\ regions for future analysis. 
\begin{table*}
  \caption{Catalogue of the 179 \hii\ regions identified in the double-ringed AM\,0644-741 galaxy and determined parameters.}
  \label{tab:results}
\begin{tabular}{ccccccccccc}
\hline
ID & \multicolumn{2}{c}{Coordinates (J2000)} &  \multicolumn{3}{c}{H$\beta$}   &  &  & kinematics     & H97 & Abundances \\
\cline{2-3}
\cline{4-6}
\cline{9-9}
\cline{11-11}

\# & R.~A.       & Dec.          & log(F(H$\beta$))         & SNR & EW    & \Av & N$_\text{O7V}$ & $\text{V}_{\text{rad}}$ & ID   & $12+\log\rm{(\frac{O}{H})}$ \\
   & [hh:mm:ss]  & [dd:mm:ss]    & [erg~cm$^{-2}$~s$^{-1}$] &     & [\AA] & [mag]   &                & [\kms]    &      &     \\
(1)& (2)         & (3)           & (4)                      & (5) & (6)   & (7)     & (8)            & (9)       & (10) & (11)\\
\hline
1  & 06:43:05.27 & --74:15:16.46 & --14.957 $\pm$ 0.072 & 13.8 & 19.4 & 0.93 $\pm$ 0.12 &  270  & 6469.3 $\pm$  6.3 & H24   & 8.50 $\pm$ 0.71 \\
2  & 06:43:04.50 & --74:15:16.10 & --15.164 $\pm$ 0.111 &  9.0 &  5.0 & 0.86 $\pm$ 0.16 &  168  & 6413.2 $\pm$ 29.0 & \dots & 8.51 $\pm$ 0.90 \\
3  & 06:43:03.70 & --74:15:15.61 & --14.728 $\pm$ 0.044 & 22.9 & 21.3 & 0.69 $\pm$ 0.08 &  457  & 6419.6 $\pm$ 16.3 & \dots & 8.51 $\pm$ 0.60 \\
4  & 06:43:03.13 & --74:15:19.27 & --15.780 $\pm$ 0.227 &  4.4 & 20.6 & 0.09 $\pm$ 0.35 &   41  & 6260.7 $\pm$  9.3 & \dots & \dots \\
5  & 06:43:03.13 & --74:15:15.67 & --14.431 $\pm$ 0.037 & 26.8 & 13.9 & 1.08 $\pm$ 0.07 &  906  & 6365.0 $\pm$ 12.1 & H26   & 8.51 $\pm$ 0.40 \\
6  & 06:43:03.31 & --74:15:13.46 & --15.032 $\pm$ 0.079 & 12.7 & 25.2 & 1.03 $\pm$ 0.14 &  227  & 6427.8 $\pm$ 10.4 & \dots & 8.53 $\pm$ 0.99 \\
7  & 06:43:03.88 & --74:15:00.47 &          \dots       &      &      &      \dots      & \dots &        $\pm$      & \dots & \dots \\
8  & 06:43:02.45 & --74:15:17.31 & --15.298 $\pm$ 0.139 &  7.2 & 16.5 & 0.82 $\pm$ 0.20 &  123  & 6287.9 $\pm$  5.2 & H26   & 8.49 $\pm$ 0.92 \\
9  & 06:43:02.57 & --74:15:14.22 & --14.137 $\pm$ 0.037 & 26.9 & 23.3 & 1.62 $\pm$ 0.08 & 1783  & 6385.0 $\pm$  6.4 & H26   & 8.52 $\pm$ 0.72 \\
10 & 06:43:01.77 & --74:15:20.53 &          \dots       &      &      &      \dots      & \dots &        $\pm$      & H27   & \dots \\
11 & 06:43:01.92 & --74:15:15.73 & --14.917 $\pm$ 0.087 & 11.5 & 20.7 & 1.17 $\pm$ 0.15 &  296  & 6307.1 $\pm$ 21.0 & \dots & 8.49 $\pm$ 0.89 \\
12 & 06:43:02.00 & --74:15:12.21 & --14.534 $\pm$ 0.053 & 18.8 & 17.0 & 1.45 $\pm$ 0.10 &  715  & 6406.7 $\pm$  7.1 & \dots & \dots \\
13 & 06:43:01.57 & --74:15:14.22 &          \dots       &      &      &      \dots      & \dots &        $\pm$      & \dots & \dots \\
14 & 06:43:01.70 & --74:15:08.94 & --15.046 $\pm$ 0.060 & 16.8 & 14.1 & 0.75 $\pm$ 0.10 &  220  & 6402.8 $\pm$  7.5 & \dots & 8.52 $\pm$ 0.75 \\
15 & 06:43:01.19 & --74:15:12.36 & --15.503 $\pm$ 0.101 &  9.9 & 12.7 & 0.00 $\pm$ 0.00 &   77  & 6325.8 $\pm$  5.7 & \dots & 8.61 $\pm$ 0.44 \\
16 & 06:43:01.14 & --74:15:10.55 & --15.020 $\pm$ 0.067 & 14.9 &  9.2 & 0.57 $\pm$ 0.10 &  233  & 6332.1 $\pm$  8.8 & \dots & 8.51 $\pm$ 0.58 \\
17 & 06:43:00.96 & --74:15:07.40 & --14.501 $\pm$ 0.036 & 27.4 & 15.3 & 0.97 $\pm$ 0.07 &  771  & 6371.3 $\pm$  3.5 & H29   & 8.53 $\pm$ 0.45 \\
18 & 06:43:00.35 & --74:15:08.07 & --14.883 $\pm$ 0.085 & 11.8 &  7.7 & 1.16 $\pm$ 0.13 &  320  & 6315.3 $\pm$  8.3 & \dots & 8.51 $\pm$ 0.98 \\
19 & 06:43:00.39 & --74:15:06.18 & --14.637 $\pm$ 0.051 & 19.7 &  9.5 & 1.21 $\pm$ 0.08 &  564  & 6301.6 $\pm$  1.1 & H30   & 8.52 $\pm$ 0.60 \\
20 & 06:43:00.28 & --74:15:05.01 & --14.951 $\pm$ 0.083 & 12.0 &  2.0 & 0.02 $\pm$ 0.10 &  274  & 6309.7 $\pm$ 13.6 & H30   & 8.53 $\pm$ 0.87 \\
21 & 06:43:00.00 & --74:15:09.72 & --15.352 $\pm$ 0.172 &  5.8 & 12.7 & 0.77 $\pm$ 0.33 &  109  & 6286.7 $\pm$  2.4 & \dots & 8.50 $\pm$ 0.89 \\
22 & 06:42:59.94 & --74:15:06.34 & --14.682 $\pm$ 0.102 &  9.8 &  5.0 & 1.79 $\pm$ 0.16 &  508  & 6313.4 $\pm$ 11.1 & H30   & 8.53 $\pm$ 0.91 \\
23 & 06:42:59.73 & --74:15:16.43 &          \dots       &      &      &      \dots      & \dots &        $\pm$      & H28   & \dots \\
24 & 06:42:59.45 & --74:15:03.06 & --14.872 $\pm$ 0.074 & 13.5 &  5.4 & 1.08 $\pm$ 0.11 &  328  & 6299.9 $\pm$ 11.2 & H31   & 8.52 $\pm$ 0.89 \\
25 & 06:42:59.97 & --74:15:00.76 & --14.853 $\pm$ 0.048 & 20.9 & 13.6 & 0.69 $\pm$ 0.08 &  343  & 6315.4 $\pm$ 14.4 & \dots & 8.50 $\pm$ 0.52 \\
26 & 06:42:59.32 & --74:15:01.29 & --14.217 $\pm$ 0.041 & 24.4 & 11.8 & 1.70 $\pm$ 0.08 & 1483  & 6298.9 $\pm$  5.5 & H31   & 8.53 $\pm$ 0.50 \\
27 & 06:42:58.97 & --74:15:00.04 & --14.136 $\pm$ 0.023 & 43.8 & 24.0 & 1.43 $\pm$ 0.04 & 1787  & 6304.7 $\pm$  5.7 & H31   & 8.53 $\pm$ 0.47 \\
28 & 06:42:59.54 & --74:14:58.08 & --14.753 $\pm$ 0.036 & 27.7 &  9.9 & 0.97 $\pm$ 0.06 &  432  & 6321.4 $\pm$  5.9 & H32   & 8.51 $\pm$ 0.58 \\
29 & 06:42:58.93 & --74:14:57.20 & --14.198 $\pm$ 0.018 & 54.8 & 25.1 & 1.09 $\pm$ 0.03 & 1550  & 6301.4 $\pm$  3.8 & H32   & 8.53 $\pm$ 0.39 \\
30 & 06:42:58.33 & --74:14:54.30 & --14.731 $\pm$ 0.036 & 27.7 & 24.0 & 0.94 $\pm$ 0.06 &  454  & 6315.3 $\pm$  4.4 & H34   & 8.51 $\pm$ 0.48 \\
31 & 06:42:59.04 & --74:14:54.54 & --14.296 $\pm$ 0.021 & 47.0 & 20.7 & 1.00 $\pm$ 0.04 & 1237  & 6333.8 $\pm$  2.7 & H33   & 8.51 $\pm$ 0.47 \\
32 & 06:42:59.57 & --74:14:54.84 & --13.986 $\pm$ 0.012 & 81.1 & 59.0 & 1.12 $\pm$ 0.02 & 2525  & 6341.9 $\pm$  2.4 & H33   & 8.51 $\pm$ 0.40 \\
33 & 06:42:59.54 & --74:14:52.82 & --14.182 $\pm$ 0.017 & 58.1 & 37.3 & 1.09 $\pm$ 0.03 & 1608  & 6334.0 $\pm$  2.2 & H33   & 8.52 $\pm$ 0.36 \\
34 & 06:42:58.70 & --74:14:52.67 & --14.231 $\pm$ 0.021 & 47.8 & 16.8 & 1.12 $\pm$ 0.04 & 1436  & 6333.5 $\pm$  3.1 & H34   & 8.52 $\pm$ 0.46 \\
35 & 06:42:57.50 & --74:14:49.45 & --15.870 $\pm$ 0.238 &  4.2 & 20.7 & 0.00 $\pm$ 0.42 &   33  & 6410.9 $\pm$ 20.3 & \dots & \dots \\
36 & 06:42:58.02 & --74:14:48.94 & --15.010 $\pm$ 0.083 & 12.0 & 14.5 & 1.19 $\pm$ 0.15 &  239  & 6347.8 $\pm$  3.1 & H35   & 8.54 $\pm$ 1.11 \\
37 & 06:42:58.56 & --74:14:48.91 & --13.872 $\pm$ 0.015 & 65.7 & 33.9 & 1.46 $\pm$ 0.03 & 3283  & 6335.9 $\pm$  5.1 & H35   & 8.56 $\pm$ 0.39 \\
38 & 06:42:57.99 & --74:14:44.31 & --14.653 $\pm$ 0.040 & 24.7 & 34.8 & 0.89 $\pm$ 0.08 &  544  & 6368.2 $\pm$  6.3 & H36   & 8.55 $\pm$ 0.65 \\
39 & 06:42:58.50 & --74:14:44.69 & --13.567 $\pm$ 0.007 &141.2 & 55.7 & 1.23 $\pm$ 0.01 & 6626  & 6349.0 $\pm$  4.0 & H36   & 8.57 $\pm$ 0.17 \\
40 & 06:42:59.16 & --74:14:44.93 & --14.957 $\pm$ 0.043 & 23.1 & 13.4 & 0.73 $\pm$ 0.08 &  270  & 6334.0 $\pm$  1.8 & \dots & 8.54 $\pm$ 0.51 \\
41 & 06:42:57.94 & --74:14:42.32 & --14.773 $\pm$ 0.051 & 19.5 & 33.3 & 1.03 $\pm$ 0.07 &  412  & 6370.9 $\pm$  9.6 & \dots & 8.53 $\pm$ 0.62 \\
42 & 06:42:58.53 & --74:14:41.14 & --14.122 $\pm$ 0.016 & 63.1 & 29.8 & 1.45 $\pm$ 0.03 & 1846  & 6371.1 $\pm$  4.8 & \dots & 8.54 $\pm$ 0.34 \\
43 & 06:42:57.80 & --74:14:39.82 & --14.872 $\pm$ 0.097 & 10.3 & 24.7 & 1.56 $\pm$ 0.16 &  328  & 6389.1 $\pm$  3.6 & \dots & 8.53 $\pm$ 0.87 \\
44 & 06:42:58.73 & --74:14:38.07 & --14.433 $\pm$ 0.020 & 50.5 & 25.1 & 0.79 $\pm$ 0.04 &  902  & 6387.1 $\pm$  3.7 & H37   & 8.52 $\pm$ 0.30 \\
45 & 06:42:58.64 & --74:14:36.73 & --14.517 $\pm$ 0.023 & 42.8 & 19.3 & 0.86 $\pm$ 0.05 &  743  & 6402.5 $\pm$  4.3 & H37   & 8.54 $\pm$ 0.40 \\
46 & 06:42:59.26 & --74:14:36.89 & --15.374 $\pm$ 0.085 & 11.7 &  7.4 & 0.53 $\pm$ 0.13 &  103  & 6371.5 $\pm$  6.1 & H37   & 8.49 $\pm$ 1.19 \\
47 & 06:42:57.66 & --74:14:34.98 & --15.510 $\pm$ 0.303 &  3.3 &  5.9 & 1.17 $\pm$ 0.46 &   76  & 6414.4 $\pm$ 17.0 & \dots & \dots \\
48 & 06:42:58.68 & --74:14:34.12 & --15.171 $\pm$ 0.074 & 13.5 &  8.4 & 1.02 $\pm$ 0.11 &  165  & 6393.5 $\pm$  5.2 & \dots & 8.54 $\pm$ 1.04 \\
49 & 06:42:57.71 & --74:14:33.88 & --15.841 $\pm$ 0.233 &  4.3 &  3.2 & 0.20 $\pm$ 0.24 &   35  & 6415.8 $\pm$  8.1 & \dots & \dots \\
50 & 06:42:58.65 & --74:14:31.19 & --15.280 $\pm$ 0.095 & 10.5 &  7.2 & 0.85 $\pm$ 0.15 &  128  & 6420.5 $\pm$ 14.2 & \dots & 8.55 $\pm$ 1.07 \\
51 & 06:42:57.71 & --74:14:30.30 & --15.321 $\pm$ 0.115 &  8.7 &  9.5 & 0.84 $\pm$ 0.17 &  117  & 6426.2 $\pm$  7.1 & \dots & 8.52 $\pm$ 1.08 \\
52 & 06:42:58.65 & --74:14:28.88 & --14.942 $\pm$ 0.123 &  8.1 &  8.5 & 1.76 $\pm$ 0.22 &  279  & 6418.8 $\pm$ 10.9 & \dots & 8.54 $\pm$ 0.88 \\
53 & 06:43:00.33 & --74:14:28.41 & --15.712 $\pm$ 0.139 &  7.2 &  5.1 & 0.47 $\pm$ 0.19 &   47  & 6444.9 $\pm$  8.3 & \dots & 8.55 $\pm$ 1.31 \\
54 & 06:42:58.86 & --74:14:27.02 & --14.983 $\pm$ 0.075 & 13.4 &  8.5 & 1.34 $\pm$ 0.14 &  254  & 6412.3 $\pm$  9.9 & H38   & 8.52 $\pm$ 0.74 \\
55 & 06:42:58.90 & --74:14:24.82 & --14.686 $\pm$ 0.048 & 20.8 & 17.2 & 1.32 $\pm$ 0.09 &  504  & 6443.3 $\pm$  5.4 & H38   & 8.54 $\pm$ 0.66 \\
56 & 06:42:59.25 & --74:14:22.03 & --15.044 $\pm$ 0.052 & 19.1 & 13.5 & 0.90 $\pm$ 0.09 &  221  & 6464.2 $\pm$  2.9 & \dots & 8.53 $\pm$ 0.98 \\
57 & 06:42:59.54 & --74:14:19.66 & --15.159 $\pm$ 0.072 & 13.8 & 11.0 & 1.03 $\pm$ 0.12 &  170  & 6453.3 $\pm$  2.9 & \dots & 8.53 $\pm$ 0.88 \\
58 & 06:43:00.94 & --74:14:18.67 & --15.709 $\pm$ 0.123 &  8.1 &  5.6 & 0.45 $\pm$ 0.16 &   48  & 6484.1 $\pm$  5.9 & \dots & \dots \\
59 & 06:42:58.33 & --74:14:18.06 & --15.601 $\pm$ 0.256 &  3.9 & 73.7 & 0.19 $\pm$ 0.69 &   61  & 6619.9 $\pm$ 10.8 & H39   & \dots \\
60 & 06:42:59.45 & --74:14:17.83 & --15.501 $\pm$ 0.072 & 13.8 &  8.7 & 0.24 $\pm$ 0.11 &   77  & 6472.1 $\pm$  4.4 & \dots & 8.52 $\pm$ 1.33 \\
\hline
\end{tabular}\\
(1) \hii\ region ID in this work;
(2-3) location: right ascension (R.A.) and declination (Dec.) (J2000);
(4) Flux of the \hb\ emission line log(F(H$\beta$)) [erg~cm$^{-2}$~s$^{-1}$];
(5) signal-to-noise ratio of \hb;
(6) equivalent width of \hb\ [\AA];
(7) Visual extinction [mag];
(8) derived number of O-type stars;
(9) radial velocity [\kms];
(10) cross ID with \citet[][]{Higdon1997};
(11) O abundances with the SLM.
\end{table*}

\begin{table*}
\footnotesize\addtolength{\tabcolsep}{-2.0pt}
  \contcaption{Catalog or \hii\ complexes in AM\,0644-741}
  \label{tab:results2}
\begin{tabular}{cccccrrrclc}
\hline
ID & \multicolumn{2}{c}{Coordinates (J2000)} &  \multicolumn{3}{c}{H$\beta$}   &  &  & kinematics     & H97 & Abundances \\
\cline{2-3}
\cline{4-6}
\cline{9-9}
\cline{11-11}

\# & R.~A.       & Dec.          & log(F(H$\beta$))         & SNR & EW    & \Av & N$_\text{O7V}$ & $\text{V}_{\text{rad}}$ & ID   & $12+\log\rm{(\frac{O}{H})}$ \\
   & [hh:mm:ss]  & [dd:mm:ss]    & [erg~cm$^{-2}$~s$^{-1}$] &     & [\AA] & [mag]   &                & [\kms]    &      &     \\
(1)& (2)         & (3)           & (4)                      & (5) & (6)   & (7)     & (8)            & (9)       & (10) & (11)\\
\hline
61  & 06:42:58.28 & --74:14:16.36 &          $\pm$       &      &      &      $\pm$      &\dots&        $\pm$      & H39   & \dots\\
62  & 06:42:59.23 & --74:14:14.76 & --14.839 $\pm$ 0.093 & 10.7 & 24.3 & 1.67 $\pm$ 0.20 & 354 & 6508.8 $\pm$ 15.1 & H40   & 8.57 $\pm$ 0.99 \\
63  & 06:42:58.86 & --74:14:14.09 & --15.241 $\pm$ 0.169 &  5.9 & 15.5 & 1.12 $\pm$ 0.34 & 140 & 6516.7 $\pm$ 19.2 & H40   & 8.54 $\pm$ 1.11 \\
64  & 06:42:59.68 & --74:14:12.13 & --14.802 $\pm$ 0.061 & 16.5 & 30.4 & 1.26 $\pm$ 0.12 & 386 & 6509.5 $\pm$  8.7 & H43   & 8.52 $\pm$ 0.76 \\
65  & 06:43:01.23 & --74:14:12.74 &          \dots       &      &      &      \dots      &\dots&        $\pm$      & \dots & \dots \\
66  & 06:42:59.82 & --74:14:10.67 & --14.597 $\pm$ 0.042 & 23.6 & 29.6 & 1.43 $\pm$ 0.09 & 618 & 6533.6 $\pm$ 11.6 & H43   & 8.55 $\pm$ 0.70 \\
67  & 06:43:00.56 & --74:14:10.11 & --15.432 $\pm$ 0.108 &  9.3 &  6.2 & 0.67 $\pm$ 0.13 &  90 & 6521.8 $\pm$  1.0 & \dots & 8.50 $\pm$ 1.19 \\
68  & 06:42:58.36 & --74:14:08.15 & --15.101 $\pm$ 0.333 &  3.0 & 45.6 & 1.74 $\pm$ 0.88 & 194 & 6602.1 $\pm$  3.7 & H42   & \dots \\
69  & 06:43:00.31 & --74:14:07.09 & --15.268 $\pm$ 0.092 & 10.9 & 16.0 & 0.90 $\pm$ 0.17 & 132 & 6538.8 $\pm$  4.6 & \dots & 8.50 $\pm$ 0.97 \\
70  & 06:42:58.52 & --74:14:03.95 & --15.794 $\pm$ 0.062 & 16.1 & \dots& 0.52 $\pm$ 0.15 &  39 & 6621.3 $\pm$  5.9 & H44   & \dots \\
71  & 06:42:59.00 & --74:14:00.72 & --15.882 $\pm$ 0.087 & 11.5 & \dots& 0.77 $\pm$ 0.24 &  32 & 6679.3 $\pm$ 47.4 & H45   &  \dots \\
72  & 06:43:00.57 & --74:14:03.22 & --15.113 $\pm$ 0.161 &  6.2 & 16.0 & 1.63 $\pm$ 0.33 & 188 & 6576.5 $\pm$ 11.4 & H46   & 8.54 $\pm$ 1.17 \\
73  & 06:43:01.12 & --74:14:02.78 & --15.056 $\pm$ 0.090 & 11.1 & 12.6 & 1.37 $\pm$ 0.16 & 215 & 6577.6 $\pm$  2.7 & H46   & 8.53 $\pm$ 1.01 \\
74  & 06:43:01.44 & --74:14:01.29 & --15.249 $\pm$ 0.135 &  7.4 &  9.6 & 0.91 $\pm$ 0.25 & 138 & 6613.6 $\pm$  6.6 & H46   & 8.52 $\pm$ 1.18 \\
75  & 06:43:01.50 & --74:14:00.10 & --15.344 $\pm$ 0.125 &  8.0 & 12.6 & 0.63 $\pm$ 0.24 & 111 & 6622.4 $\pm$  9.6 & H46   & \dots \\
76  & 06:43:00.81 & --74:13:58.06 & --15.575 $\pm$ 0.192 &  5.2 & 21.3 & 0.54 $\pm$ 0.34 &  65 & 6883.1 $\pm$ 25.2 & \dots & \dots \\
77  & 06:43:00.03 & --74:13:56.29 & --14.740 $\pm$ 0.097 & 10.3 &255.1 & 0.99 $\pm$ 0.20 & 445 & 6647.2 $\pm$  2.1 & H47   & \dots \\
78  & 06:43:02.26 & --74:13:57.19 & --14.598 $\pm$ 0.044 & 22.8 & 23.5 & 1.13 $\pm$ 0.08 & 617 & 6679.3 $\pm$ 21.1 & H49   & 8.54 $\pm$ 0.78 \\
79  & 06:43:02.57 & --74:13:55.50 & --14.891 $\pm$ 0.096 & 10.4 & 13.2 & 1.15 $\pm$ 0.19 & 314 & 6738.5 $\pm$ 41.9 & H49   & 8.54 $\pm$ 0.69 \\
80  & 06:43:03.52 & --74:13:55.07 & --15.506 $\pm$ 0.204 &  4.9 &  5.5 & 0.63 $\pm$ 0.33 &  76 & 6743.4 $\pm$  9.2 & \dots & \dots \\
81  & 06:43:00.77 & --74:13:53.53 & --15.550 $\pm$ 0.278 &  3.6 & 31.2 & 0.74 $\pm$ 0.55 &  69 & 6714.0 $\pm$ 10.0 & \dots & \dots \\
82  & 06:43:01.54 & --74:13:51.26 & --14.820 $\pm$ 0.092 & 10.9 &116.5 & 0.86 $\pm$ 0.18 & 370 & 6738.8 $\pm$  8.7 & H48   & 8.57 $\pm$ 0.84 \\
83  & 06:43:04.49 & --74:13:52.42 & --14.889 $\pm$ 0.051 & 19.7 & 15.9 & 0.91 $\pm$ 0.10 & 316 & 6795.9 $\pm$ 15.4 & H51   & 8.56 $\pm$ 0.79 \\
84  & 06:43:02.15 & --74:13:49.26 & --14.445 $\pm$ 0.027 & 36.9 & 96.7 & 0.71 $\pm$ 0.05 & 877 & 6745.7 $\pm$  5.6 & H50   & 8.56 $\pm$ 0.58 \\
85  & 06:43:03.65 & --74:13:51.95 & --14.967 $\pm$ 0.049 & 20.3 & 22.4 & 0.52 $\pm$ 0.09 & 264 & 6869.4 $\pm$ 16.7 & H51   & 8.49 $\pm$ 0.61 \\
86  & 06:43:03.96 & --74:13:50.38 & --14.661 $\pm$ 0.045 & 22.2 & 39.5 & 0.69 $\pm$ 0.10 & 534 & 6821.5 $\pm$ 20.4 & H51   & 8.58 $\pm$ 0.52 \\
87  & 06:43:03.08 & --74:13:47.42 & --15.318 $\pm$ 0.125 &  8.0 & 25.2 & 0.63 $\pm$ 0.24 & 118 & 6806.1 $\pm$  3.3 & \dots & 8.52 $\pm$ 1.10 \\
88  & 06:43:03.18 & --74:13:45.67 & --15.479 $\pm$ 0.172 &  5.8 & 18.4 & 0.57 $\pm$ 0.35 &  81 & 6807.5 $\pm$  4.6 & \dots & 8.51 $\pm$ 1.16 \\
89  & 06:43:03.54 & --74:13:46.33 & --15.318 $\pm$ 0.145 &  6.9 & 16.5 & 0.98 $\pm$ 0.27 & 118 & 6809.1 $\pm$  5.3 & \dots & 8.52 $\pm$ 0.88 \\
90  & 06:43:03.95 & --74:13:44.52 & --15.466 $\pm$ 0.164 &  6.1 & 18.2 & 0.57 $\pm$ 0.32 &  84 & 6828.1 $\pm$ 13.0 & \dots & 8.53 $\pm$ 0.97 \\
91  & 06:43:04.43 & --74:13:44.07 & --15.066 $\pm$ 0.092 & 10.9 & 23.0 & 0.75 $\pm$ 0.19 & 210 & 6833.0 $\pm$  8.0 & H52   & 8.54 $\pm$ 0.75 \\
92  & 06:43:04.80 & --74:13:44.17 & --14.659 $\pm$ 0.049 & 20.3 & 29.9 & 1.05 $\pm$ 0.09 & 536 & 6836.3 $\pm$  4.3 & H52   & 8.53 $\pm$ 0.65 \\
93  & 06:43:05.01 & --74:13:50.39 & --15.213 $\pm$ 0.064 & 15.6 & 17.2 & 0.34 $\pm$ 0.11 & 150 & 6869.3 $\pm$  6.5 & H53   & 8.51 $\pm$ 0.89 \\
94  & 06:43:05.48 & --74:13:43.90 & --15.068 $\pm$ 0.097 & 10.3 & 15.1 & 0.96 $\pm$ 0.17 & 209 & 6847.0 $\pm$  2.9 & \dots & 8.56 $\pm$ 0.54 \\
95  & 06:43:05.63 & --74:13:51.30 & --14.320 $\pm$ 0.015 & 65.9 & 40.6 & 0.54 $\pm$ 0.03 &1170 & 6852.5 $\pm$  6.9 & H53   & 8.55 $\pm$ 0.44 \\
96  & 06:43:05.94 & --74:13:42.63 & --15.323 $\pm$ 0.189 &  5.3 &  6.3 & 1.05 $\pm$ 0.33 & 116 & 6908.7 $\pm$ 14.4 & \dots & \dots \\
97  & 06:43:06.59 & --74:13:44.09 & --15.461 $\pm$ 0.128 &  7.8 & 14.3 & 0.27 $\pm$ 0.25 &  85 & 6975.0 $\pm$ 29.2 & \dots & 8.47 $\pm$ 0.87 \\
98  & 06:43:06.64 & --74:13:50.90 & --15.157 $\pm$ 0.068 & 14.8 & 12.0 & 0.53 $\pm$ 0.12 & 170 & 6878.0 $\pm$ 16.3 & \dots & 8.55 $\pm$ 0.86 \\
99  & 06:43:07.23 & --74:13:49.16 & --15.196 $\pm$ 0.106 &  9.4 & 13.4 & 0.63 $\pm$ 0.21 & 156 & 6913.7 $\pm$ 10.0 & \dots & 8.49 $\pm$ 0.83 \\
100 & 06:43:07.45 & --74:13:44.50 & --14.781 $\pm$ 0.036 & 27.7 & 34.6 & 0.38 $\pm$ 0.07 & 405 & 6860.6 $\pm$  6.7 & H1    & 8.51 $\pm$ 0.52 \\
101 & 06:43:07.74 & --74:13:52.11 & --14.457 $\pm$ 0.034 & 29.1 & 23.6 & 1.17 $\pm$ 0.06 & 854 & 6849.6 $\pm$  9.9 & \dots & 8.57 $\pm$ 0.94 \\
102 & 06:43:08.89 & --74:13:49.28 & --15.574 $\pm$ 0.152 &  6.6 &  5.9 & 0.29 $\pm$ 0.23 &  65 & 6934.2 $\pm$ 14.0 & \dots & 8.49 $\pm$ 0.85 \\
103 & 06:43:08.35 & --74:13:53.72 & --14.106 $\pm$ 0.010 & 98.6 & 34.3 & 0.85 $\pm$ 0.02 &1915 & 6860.8 $\pm$  5.1 & H2    & 8.57 $\pm$ 0.75 \\
104 & 06:43:09.27 & --74:13:52.38 & --15.152 $\pm$ 0.068 & 14.8 &  8.8 & 0.56 $\pm$ 0.12 & 172 & 6931.2 $\pm$ 11.5 & \dots & 8.52 $\pm$ 0.78 \\
105 & 06:43:09.26 & --74:13:53.81 & --14.945 $\pm$ 0.062 & 16.1 & 11.4 & 0.76 $\pm$ 0.12 & 277 & 6915.9 $\pm$  7.0 & \dots & 8.52 $\pm$ 0.79 \\
106 & 06:43:08.75 & --74:13:55.78 & --14.450 $\pm$ 0.036 & 27.9 & 20.8 & 1.08 $\pm$ 0.07 & 867 & 6875.5 $\pm$  4.5 & H3    & 8.56 $\pm$ 0.75 \\
107 & 06:43:09.17 & --74:13:57.76 & --13.973 $\pm$ 0.014 & 71.4 & 62.4 & 0.96 $\pm$ 0.03 &2602 & 6864.6 $\pm$  3.2 & H3    & 8.63 $\pm$ 0.42 \\
108 & 06:43:10.37 & --74:13:56.05 & --15.715 $\pm$ 0.278 &  3.6 &  3.2 & 0.46 $\pm$ 0.40 &  47 & 6948.1 $\pm$ 13.1 & \dots & \dots \\
109 & 06:43:09.64 & --74:14:01.53 & --14.344 $\pm$ 0.022 & 44.7 & 24.3 & 1.19 $\pm$ 0.05 &1107 & 6849.1 $\pm$  2.2 & H4    & 8.57 $\pm$ 0.50 \\
110 & 06:43:09.89 & --74:14:02.83 & --14.381 $\pm$ 0.020 & 50.3 & 23.9 & 0.98 $\pm$ 0.03 &1017 & 6838.6 $\pm$ 11.4 & H4    & 8.54 $\pm$ 0.49 \\
111 & 06:43:11.07 & --74:14:01.87 & --14.834 $\pm$ 0.036 & 27.9 & 19.3 & 0.86 $\pm$ 0.06 & 358 & 6863.8 $\pm$  4.2 & H5    & 8.52 $\pm$ 0.53 \\
112 & 06:43:11.27 & --74:14:04.04 & --15.175 $\pm$ 0.072 & 13.9 &  9.9 & 0.88 $\pm$ 0.11 & 163 & 6871.6 $\pm$ 18.4 & H5    & 8.53 $\pm$ 0.71 \\
113 & 06:43:10.01 & --74:14:05.05 & --14.645 $\pm$ 0.036 & 28.0 & 18.8 & 1.10 $\pm$ 0.06 & 554 & 6828.9 $\pm$ 17.9 & \dots & 8.54 $\pm$ 0.74 \\
114 & 06:43:11.36 & --74:14:08.15 & --15.393 $\pm$ 0.086 & 11.6 &  9.6 & 0.44 $\pm$ 0.15 &  99 & 6887.9 $\pm$  9.9 & H6    & 8.47 $\pm$ 0.95 \\
115 & 06:43:11.38 & --74:14:10.26 & --14.776 $\pm$ 0.037 & 27.1 & 12.3 & 0.93 $\pm$ 0.07 & 409 & 6863.2 $\pm$  3.9 & H6    & 8.50 $\pm$ 0.69 \\
116 & 06:43:10.20 & --74:14:11.63 & --15.105 $\pm$ 0.077 & 13.0 &  8.3 & 1.12 $\pm$ 0.13 & 192 & 6802.6 $\pm$ 14.1 & H7    & 8.51 $\pm$ 1.36 \\
117 & 06:43:11.26 & --74:14:12.23 & --15.134 $\pm$ 0.061 & 16.3 &  9.2 & 0.72 $\pm$ 0.10 & 180 & 6864.7 $\pm$  5.8 & H6    & 8.51 $\pm$ 0.82 \\
118 & 06:43:11.66 & --74:14:12.62 & --15.634 $\pm$ 0.164 &  6.1 &  3.1 & 0.25 $\pm$ 0.22 &  57 & 6866.3 $\pm$  3.3 & H6    & 8.47 $\pm$ 1.15 \\
119 & 06:43:13.20 & --74:14:13.88 &          \dots       &      &      &      \dots      &\dots&        $\pm$      & \dots & \dots \\
120 & 06:43:10.22 & --74:14:14.45 & --15.291 $\pm$ 0.079 & 12.7 & 11.4 & 0.59 $\pm$ 0.12 & 125 & 6785.7 $\pm$ 12.3 & H7    & 8.53 $\pm$ 0.93 \\
\hline
\end{tabular}
\end{table*}

\begin{table*}
\footnotesize\addtolength{\tabcolsep}{-2.0pt}
  \contcaption{Catalog or \hii\ complexes in AM\,0644-741}
  \label{tab:results3}
\begin{tabular}{cccccrrrclc}
\hline
ID & \multicolumn{2}{c}{Coordinates (J2000)} &  \multicolumn{3}{c}{H$\beta$}   &  &  & kinematics     & H97 & Abundances \\
\cline{2-3}
\cline{4-6}
\cline{9-9}
\cline{11-11}
\# & R.~A.       & Dec.          & log(F(H$\beta$))         & SNR & EW    & \Av & N$_\text{O7V}$ & $\text{V}_{\text{rad}}$ & ID   & $12+\log\rm{(\frac{O}{H})}$ \\
   & [hh:mm:ss]  & [dd:mm:ss]    & [erg~cm$^{-2}$~s$^{-1}$] &     & [\AA] & [mag]   &                & [\kms]    &      &     \\
(1)& (2)         & (3)           & (4)                      & (5) & (6)   & (7)     & (8)            & (9)       & (10) & (11)\\
\hline
121 & 06:43:10.85 & --74:14:16.61 & --15.052 $\pm$ 0.091 & 11.0 &  9.1 & 1.17 $\pm$ 0.16 & 217 & 6806.3 $\pm$ 34.4 & H8    & \dots  \\
122 & 06:43:11.47 & --74:14:18.25 & --14.922 $\pm$ 0.072 & 13.8 &  8.8 & 1.14 $\pm$ 0.12 & 293 & 6800.0 $\pm$  3.2 & H8    & 8.51 $\pm$ 0.87 \\
123 & 06:43:10.47 & --74:14:18.80 & --14.928 $\pm$ 0.055 & 18.3 &  9.6 & 1.11 $\pm$ 0.09 & 289 & 6753.1 $\pm$ 10.1 & \dots & 8.53 $\pm$ 1.15 \\
124 & 06:43:10.71 & --74:14:21.14 & --14.824 $\pm$ 0.046 & 21.8 & 14.5 & 1.12 $\pm$ 0.09 & 367 & 6751.8 $\pm$  2.2 & \dots & 8.53 $\pm$ 0.68 \\
125 & 06:43:11.21 & --74:14:22.14 & --15.389 $\pm$ 0.130 &  7.7 &  4.7 & 0.93 $\pm$ 0.19 & 100 & 6784.9 $\pm$  3.1 & \dots & 8.50 $\pm$ 1.11 \\
126 & 06:43:09.10 & --74:14:22.22 & --15.383 $\pm$ 0.108 &  9.3 &  6.7 & 0.80 $\pm$ 0.19 & 101 & 6629.8 $\pm$ 12.0 & \dots & 8.52 $\pm$ 1.00 \\
127 & 06:43:09.51 & --74:14:23.02 & --15.266 $\pm$ 0.070 & 14.2 & 11.0 & 0.73 $\pm$ 0.11 & 133 & 6651.6 $\pm$  6.5 & \dots & 8.54 $\pm$ 1.03 \\
128 & 06:43:10.63 & --74:14:24.05 & --15.036 $\pm$ 0.053 & 18.9 &  9.3 & 0.95 $\pm$ 0.08 & 225 & 6714.1 $\pm$  8.8 & H9    & 8.54 $\pm$ 0.74 \\
129 & 06:43:11.51 & --74:14:25.67 & --15.585 $\pm$ 0.096 & 10.4 &  2.7 & 0.39 $\pm$ 0.08 &  64 & 6802.8 $\pm$  9.7 & \dots & 8.50 $\pm$ 1.44 \\
130 & 06:43:11.39 & --74:14:28.27 & --15.427 $\pm$ 0.114 &  8.8 &  3.3 & 0.73 $\pm$ 0.12 &  91 & 6793.5 $\pm$  6.6 & \dots & 8.49 $\pm$ 1.09 \\
131 & 06:43:10.88 & --74:14:28.66 & --14.788 $\pm$ 0.034 & 29.8 & 13.1 & 0.89 $\pm$ 0.06 & 398 & 6782.9 $\pm$  1.3 & H11   & 8.51 $\pm$ 0.49 \\
132 & 06:43:10.12 & --74:14:25.82 & --14.719 $\pm$ 0.041 & 24.2 & 10.2 & 1.26 $\pm$ 0.07 & 467 & 6672.7 $\pm$  7.0 & H9    & 8.52 $\pm$ 0.83 \\
133 & 06:43:09.65 & --74:14:26.74 & --14.925 $\pm$ 0.045 & 22.3 & 11.1 & 1.00 $\pm$ 0.07 & 291 & 6632.6 $\pm$  8.6 & H9    & 8.51 $\pm$ 0.89 \\
134 & 06:43:08.93 & --74:14:27.24 & --14.351 $\pm$ 0.020 & 50.3 & 26.1 & 1.27 $\pm$ 0.04 &1090 & 6549.7 $\pm$  2.9 & H10   & 8.53 $\pm$ 0.62 \\
135 & 06:43:08.49 & --74:14:24.66 & --15.316 $\pm$ 0.094 & 10.6 &  4.3 & 0.84 $\pm$ 0.15 & 118 & 6502.0 $\pm$ 20.0 & \dots & \dots \\
136 & 06:43:08.33 & --74:14:25.13 & --15.336 $\pm$ 0.101 &  9.9 &  6.4 & 0.96 $\pm$ 0.17 & 113 & 6500.0 $\pm$ 17.8 & \dots & \dots \\
137 & 06:43:07.03 & --74:14:21.39 & --15.841 $\pm$ 0.333 &  3.0 &  0.6 & 0.00 $\pm$ 0.00 &  35 & 6356.4 $\pm$  0.0 & \dots & \dots \\
138 & 06:43:06.81 & --74:14:24.21 & --15.737 $\pm$ 0.208 &  4.8 &  0.9 & 0.00 $\pm$ 0.13 &  45 & 6349.1 $\pm$ 12.9 & \dots & \dots \\
139 & 06:43:10.65 & --74:14:30.53 & --14.556 $\pm$ 0.021 & 48.4 & 16.9 & 0.81 $\pm$ 0.04 & 680 & 6777.0 $\pm$  4.5 & H11   & 8.51 $\pm$ 0.46 \\
140 & 06:43:10.55 & --74:14:33.01 & --14.750 $\pm$ 0.028 & 36.1 & 14.6 & 0.69 $\pm$ 0.05 & 435 & 6751.7 $\pm$ 10.1 & H12   & 8.51 $\pm$ 0.53 \\
141 & 06:43:11.11 & --74:14:34.57 & --14.080 $\pm$ 0.014 & 70.1 & 34.5 & 1.22 $\pm$ 0.03 &2033 & 6768.7 $\pm$  3.3 & H12   & 8.52 $\pm$ 0.28 \\
142 & 06:43:10.46 & --74:14:35.18 & --14.437 $\pm$ 0.017 & 57.7 &  8.9 & 0.48 $\pm$ 0.03 & 894 & 6726.6 $\pm$  2.8 & H12   & 8.52 $\pm$ 0.25 \\
143 & 06:43:10.45 & --74:14:36.57 & --14.554 $\pm$ 0.028 & 36.1 &  8.7 & 0.51 $\pm$ 0.05 & 683 & 6724.7 $\pm$  4.4 & H12   & 8.52 $\pm$ 0.46 \\
144 & 06:43:09.15 & --74:14:36.20 & --15.564 $\pm$ 0.145 &  6.9 &  4.2 & 0.80 $\pm$ 0.21 &  67 & 6638.5 $\pm$  7.5 & \dots & 8.55 $\pm$ 0.84 \\
145 & 06:43:10.29 & --74:14:38.75 & --15.032 $\pm$ 0.050 & 20.2 &  5.7 & 0.65 $\pm$ 0.06 & 227 & 6726.5 $\pm$  3.1 & \dots & 8.52 $\pm$ 0.78 \\
146 & 06:43:09.59 & --74:14:39.18 & --14.943 $\pm$ 0.061 & 16.4 &  7.7 & 1.19 $\pm$ 0.09 & 279 & 6671.7 $\pm$  6.6 & \dots & 8.51 $\pm$ 0.97 \\
147 & 06:43:10.62 & --74:14:40.50 & --15.291 $\pm$ 0.084 & 11.9 &  7.1 & 0.69 $\pm$ 0.11 & 125 & 6739.5 $\pm$  3.9 & \dots & 8.52 $\pm$ 0.82 \\
148 & 06:43:11.23 & --74:14:42.04 & --14.994 $\pm$ 0.046 & 21.7 & 24.4 & 0.75 $\pm$ 0.07 & 248 & 6825.6 $\pm$  2.5 & H13   & 8.48 $\pm$ 0.64 \\
149 & 06:43:09.57 & --74:14:42.33 & --15.290 $\pm$ 0.078 & 12.8 &  6.0 & 0.80 $\pm$ 0.08 & 125 & 6686.1 $\pm$  5.3 & \dots & 8.51 $\pm$ 1.31 \\
150 & 06:43:08.99 & --74:14:42.24 & --15.178 $\pm$ 0.083 & 12.1 & 22.0 & 0.89 $\pm$ 0.13 & 162 & 6638.0 $\pm$  0.7 & \dots & 8.56 $\pm$ 1.14 \\
151 & 06:43:10.38 & --74:14:44.26 & --15.211 $\pm$ 0.056 & 17.8 & 13.4 & 0.34 $\pm$ 0.10 & 150 & 6715.3 $\pm$  3.3 & H14   & 8.52 $\pm$ 0.43 \\
152 & 06:43:10.08 & --74:14:45.78 & --14.701 $\pm$ 0.025 & 39.4 & 20.6 & 0.54 $\pm$ 0.04 & 487 & 6703.9 $\pm$  3.5 & H14   & 8.52 $\pm$ 0.36 \\
153 & 06:43:10.87 & --74:14:48.93 & --15.469 $\pm$ 0.161 &  6.2 & 16.2 & 0.70 $\pm$ 0.29 &  83 & 6767.9 $\pm$  4.1 & \dots & 8.46 $\pm$ 1.04 \\
154 & 06:43:10.07 & --74:14:48.56 & --15.321 $\pm$ 0.071 & 14.1 &  4.7 & 0.33 $\pm$ 0.10 & 117 & 6716.6 $\pm$  4.6 & H16   & 8.52 $\pm$ 0.68 \\
155 & 06:43:09.90 & --74:14:50.19 & --15.251 $\pm$ 0.069 & 14.5 &  7.5 & 0.44 $\pm$ 0.12 & 137 & 6712.7 $\pm$ 14.3 & H16   & 8.50 $\pm$ 0.74 \\
156 & 06:43:09.46 & --74:14:50.27 & --15.161 $\pm$ 0.072 & 13.8 &  9.2 & 0.67 $\pm$ 0.12 & 169 & 6676.4 $\pm$ 12.3 & H16   & 8.51 $\pm$ 0.73 \\
157 & 06:43:09.14 & --74:14:51.11 & --15.411 $\pm$ 0.122 &  8.2 & 11.7 & 0.63 $\pm$ 0.23 &  95 & 6666.8 $\pm$  1.2 & \dots & 8.54 $\pm$ 0.93 \\
158 & 06:43:07.01 & --74:14:49.01 &          \dots       &      &      &      \dots      &\dots&        $\pm$      & \dots & \dots \\
159 & 06:43:10.42 & --74:14:52.64 & --15.705 $\pm$ 0.196 &  5.1 & 48.2 & 0.39 $\pm$ 0.33 &  48 & 6732.7 $\pm$  9.9 & \dots & \dots \\
160 & 06:43:11.05 & --74:14:54.43 & --15.771 $\pm$ 0.068 & 14.8 &\dots & 1.00 $\pm$ 0.17 &  41 & 6790.4 $\pm$  3.0 & H17   & \dots \\
161 & 06:43:11.93 & --74:14:54.83 & --15.868 $\pm$ 0.083 & 12.1 &\dots & 0.62 $\pm$ 0.25 &  33 & 6839.9 $\pm$ 21.5 & H15   &  \dots \\
162 & 06:43:09.31 & --74:14:54.65 & --14.945 $\pm$ 0.049 & 20.6 &  9.7 & 0.78 $\pm$ 0.08 & 277 & 6673.6 $\pm$  3.4 & H18   & 8.50 $\pm$ 0.69 \\
163 & 06:43:08.90 & --74:14:55.48 & --15.356 $\pm$ 0.093 & 10.8 &  6.8 & 0.61 $\pm$ 0.15 & 108 & 6652.9 $\pm$  9.2 & H18   & 8.52 $\pm$ 0.82 \\
164 & 06:43:10.59 & --74:14:56.96 & --15.787 $\pm$ 0.037 & 27.3 &\dots & 0.57 $\pm$ 0.09 &  40 & 6784.8 $\pm$ 13.6 & H17   & \dots \\
165 & 06:43:11.62 & --74:14:57.36 & --15.797 $\pm$ 0.111 &  9.0 &\dots & 1.42 $\pm$ 0.32 &  39 & 6822.5 $\pm$ 67.8 & \dots & \dots \\
166 & 06:43:10.32 & --74:14:58.35 & --15.947 $\pm$ 0.046 & 21.9 &\dots & 0.40 $\pm$ 0.10 &  28 & 6769.8 $\pm$ 18.8 & H19   & \dots \\
167 & 06:43:09.21 & --74:14:58.00 & --15.703 $\pm$ 0.147 &  6.8 & 10.2 & 0.39 $\pm$ 0.21 &  48 & 6680.2 $\pm$  8.1 & \dots & \dots \\
168 & 06:43:08.71 & --74:14:58.70 & --15.927 $\pm$ 0.182 &  5.5 &  5.5 & 0.18 $\pm$ 0.25 &  29 & 6622.8 $\pm$ 14.9 & \dots & 8.52 $\pm$ 0.87 \\
169 & 06:43:10.23 & --74:15:01.30 & --15.507 $\pm$ 0.175 &  5.7 &\dots & 2.57 $\pm$ 0.54 &  76 & 6724.4 $\pm$  5.9 & H19   & \dots \\
170 & 06:43:07.46 & --74:15:03.44 & --16.011 $\pm$ 0.313 &  3.2 &  3.2 & 0.00 $\pm$ 0.42 &  24 & 6586.9 $\pm$ 22.4 & H21   & \dots \\
171 & 06:43:07.91 & --74:15:05.86 &          \dots       &      &      &      \dots      &\dots&        $\pm$      & H21   & \dots \\
172 & 06:43:09.05 & --74:15:06.92 & --14.607 $\pm$ 0.026 & 38.4 & 24.2 & 0.56 $\pm$ 0.03 & 604 & 6680.3 $\pm$  7.3 & H20   & 8.49 $\pm$ 0.71 \\
173 & 06:43:09.73 & --74:15:06.99 & --15.284 $\pm$ 0.097 & 10.3 & 35.7 & 0.33 $\pm$ 0.12 & 127 & 6679.4 $\pm$ 10.4 & H20   & 8.49 $\pm$ 0.71 \\
174 & 06:43:09.41 & --74:15:08.33 & --14.885 $\pm$ 0.032 & 30.9 & 29.0 & 0.31 $\pm$ 0.04 & 319 & 6682.4 $\pm$  1.5 & H20   & 8.49 $\pm$ 0.70 \\
175 & 06:43:09.00 & --74:15:09.43 & --15.273 $\pm$ 0.079 & 12.7 & 23.6 & 0.16 $\pm$ 0.14 & 130 & 6671.9 $\pm$  6.4 & H20   & 8.48 $\pm$ 0.58 \\
176 & 06:43:08.61 & --74:15:10.78 & --15.542 $\pm$ 0.213 &  4.7 & 63.3 & 0.38 $\pm$ 0.33 &  70 & 6650.9 $\pm$  8.5 & \dots & \dots \\
177 & 06:43:08.69 & --74:15:12.43 & --15.692 $\pm$ 0.323 &  3.1 & 28.9 & 0.61 $\pm$ 0.36 &  50 & 6638.4 $\pm$ 13.7 & \dots & \dots \\
178 & 06:43:06.45 & --74:15:07.96 & --15.543 $\pm$ 0.196 &  5.1 &  6.2 & 0.76 $\pm$ 0.28 &  70 & 6520.4 $\pm$  3.5 & H22   & \dots \\
179 & 06:43:05.64 & --74:15:09.09 & --15.036 $\pm$ 0.227 &  4.4 & 13.9 & 1.86 $\pm$ 0.45 & 225 & 6476.2 $\pm$ 16.7 & H23   & 8.51 $\pm$ 1.35 \\
\hline
\end{tabular}
\end{table*}

\section{AGN confirmation}\label{AGN confirmation}

\begin{figure*}
\begin{center}
\includegraphics[width=1\linewidth]{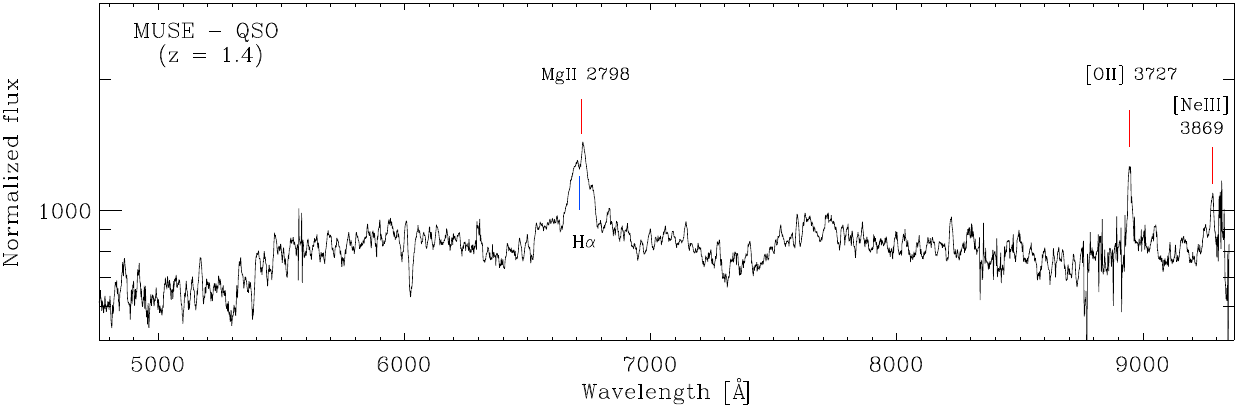}
\caption{MUSE spectrum of a 'false' \hii\ region candidate in the AM\,0644-741 galaxy.
Three emission lines are identified at z$\sim$1.4 and are indicated above the continuum (red line).
The false \ha\ corresponds to \mgiia\ \citep[][]{Heida+2013}.
Although, on top of the wide emission line we do identified the narrow \ha\ line in absorption
corresponding to the redshift of AM\,0644-741 (z$\sim$0.022).
Additionally to \citet[][]{Heida+2013} we identified \oiia\ and \neiiia.
Flux is shown normalized to 10$^{-20}$ erg\,cm$^{-2}$\,s$^{-1}$\,\AA$^{-1}$.}
\label{fig:qso}
\end{center}
\end{figure*}

In Fig.~\ref{fig:qso} we show the MUSE spectrum of a "false" \hii\ region candidate in the AM\,0644-741 galaxy.
The location of this  QSO is indicated in Fig.~\ref{fig:muse_image} by a red circle labeled as the X-ray source W3 \citep[][]{Wolter+2018}.
The spectrum contains information from astronomical objects at different distances, such as the \ha\ line in absorption at the redshift of AM\,0644-741.
We detected two emission lines of \oiia\ and \neiiia\ at z$\sim$1.4, confirming its AGN nature,
as first reported by \citet[][]{Heida+2013}.

\end{document}